\documentclass[preprint]{aastex}
\usepackage{natbib}
\usepackage[caption=false]{subfig}
\usepackage{graphicx}
\usepackage{acronym}
\usepackage{hyperref}
\usepackage{bm}
\usepackage{epstopdf}
\usepackage{multirow}
\graphicspath{{images/}}
\begin{document}
\title{Optimizing Global Coronal Magnetic Field Models Using Image-Based Constraints}
\author{Shaela I.\ Jones\altaffilmark{1}\email{shaela.i.jones-mecholsky@nasa.gov} \and Joseph M.\ Davila \and Vadim Uritsky \altaffilmark{1}}
\affil{NASA Goddard Space Flight Center, Code 670, Greenbelt, MD  20771}
\altaffiltext{1}{Catholic University of America}
\email{shaela.i.jonesmecholsky@nasa.gov}

\begin{abstract}
The coronal magnetic field directly or indirectly affects a majority of the phenomena studied in space physics.  It provides energy for coronal heating, controls the release of \acp{cme}, and drives heliospheric and magnetospheric activitiy, yet the coronal magnetic field itself has proven difficult to measure.  This difficulty has prompted a decades-long effort to develop accurate, timely, models of the field - an effort that continues today.  We have developed a method for improving global coronal magnetic field models by incorporating the type of morphological constraints which could be derived from coronal images.  Here we report promising initial tests of this approach on two theoretical problems, and discuss opportunities for application.
\end{abstract}
\acrodef{pfss}[PFSS]{Potential Field Source Surface}
\acrodef{euvi}[EUVI]{Extreme Ultraviolet Imager}
\acrodef{pos}[POS]{plane of the sky}
\acrodef{pb}[pB]{polarized brightness}
\acrodef{stereo}[STEREO]{Solar TErrestrial RElations Observatory}
\acrodef{cme}[CME]{coronal mass ejection}
\acrodef{3d}[3D]{three-dimensional}
\acrodef{los}[LOS]{line-of-sight}
\acrodef{imf}[IMF]{interplanetary magnetic field}
\acrodef{nlfff}[NLFFF]{Non-Linear Force Free Field}
\acrodef{lfff}[LFFF]{Linear Force Free Field}
\acrodef{rms}[RMS]{root-mean-square}
\acrodef{euv}[EUV]{extreme ultraviolet}
\acrodef{mae}[MAE]{mean absolute error}
\acrodef{sht}[SHT]{spherical harmonic transform}
\acrodef{gong}[GONG]{global oscillation network group}
\acrodef{mhd}[MHD]{magnetohydrodynamics}
\acrodef{spp}[SPP]{Solar Probe Plus}
\acrodef{so}[SO]{Solar Orbiter}
\acrodef{mlso}[MLSO]{Mauna Loa Solar Observatory}
\section{Introduction}
\label{introduction}
In this paper we will present a method we have developed to increase the accuracy of coronal magnetic field models using morphological information of the sort that can be derived from coronal images.  The determination of the coronal magnetic field is a critical and long-standing problem for space physics.  The coronal magnetic field determines the plasma dynamics not only in the corona but also, indirectly, the \ac{imf} and the planetary magnetospheres.  The coronal magnetic field must be known in order to accurately predict \ac{cme} trajectories \citep{kay13,panasenco13} and solar wind conditions \citep{schrijver03,neugebauer02}, including conditions at Earth.  However, the coronal magnetic field is difficult to measure.  Some techniques, such as spectro-polarimetry and radiographic measurements, are able to determine the coronal magnetic field in select regions, but it is still not possible to routinely measure the magnetic field throughout the majority of the corona.  These techniques tend to work best in regions with high emission rates and/or high field strength.  Even where measurements can be made, they are often subject to interpretation due the optically thin nature of the corona \citep{cargill09}.

In the absence of reliable measurements, researchers have developed models that can be used to extrapolate the photospheric magnetic field into the corona and beyond.  The three most commonly used field models (excepting solution of the full \ac{mhd} equations) assume that current in the corona is strictly parallel to the magnetic field, resulting in an equation of the form:
\begin{equation}
\label{eqn_potential}
\boldsymbol{\nabla} \times \boldsymbol{B} = \alpha(\boldsymbol{r}) \boldsymbol{B}
\end{equation}
where $\boldsymbol{B}$ is the magnetic field and $\alpha(\boldsymbol{r})$ is a function (or constant) to be determined by the modeller.

One of the earliest of these models, the \ac{pfss} model \citep{schatten69, altschuler69}, uses $\alpha=0$ everywhere, resulting in a scalar magnetic potential satisfying Laplace's equation \citep{mackay12}.  To solve Laplace's equation, one assumes that the magnetic field becomes entirely radial at a spherically shaped upper boundary (the so-called source surface) in the corona, generally taken to be $2.5 R_\sun$ from Sun center (though the optimal source surface height may vary - see works by \citet{arden14,lee11}).   Synoptic photospheric magnetograms are then taken as the lower boundary condition.  The \ac{pfss} model remains in frequent use today because it is straightforward to understand, quick to compute, and has a unique solution.  However, the condition that $\alpha=0$ means that there are no currents anywhere in the volume.  This is troubling given that the model region includes the chromosphere.  Furthermore, a potential field is the minimum-energy field configuration, which means that it cannot be used to accurately model regions of coronal activity, where stored energy is being released and driving some of the most interesting phenomena.

More sophisticated models, such as the \ac{lfff} (uniform $\alpha(\boldsymbol{r})=\alpha$) and \ac{nlfff} ($\alpha(\boldsymbol{r})$ spatially varying) \citep{gary89,metcalf08} are generally applied only to small areas such as active regions (though see \citet{tadesse14b}).  Further, different methods of handling boundary conditions and calculating the models can result in variable field configurations \citep{derosa09,schrijver06}.  Additionally, even the \ac{lfff} and \ac{nlfff} models cannot be said to be realistic in the chromosphere, where in some places the plasma $\beta$ approaches one \citep{gary01} and equation \ref{eqn_potential} may be untrue for a significant fraction of the model volume.  

\ac{mhd} models are more physically realistic and include information about the plasma.  Also, \ac{mhd} models can directly model time-dependent phenomena (e.g. \citet{devore08, antiochos11, edmondson10}.)  But they suffer from some of the same problems as the \ac{nlfff} models in terms of non-uniqueness and computational demand \citep{vasquez08,mackay12}.  Often in \ac{mhd} models trade-offs must be made between computational demand and the size and resolution of the simulated region \citep{feng12}, or the physical treatment of the photosphere/corona transition.  The nature of these trade-offs makes different \ac{mhd} models preferable in different situations \citep{riley11}.  (For detailed reviews of current coronal magnetic field modelling methods, see \emph{e.g.} \citet{regnier13} and \citet{mackay12}.)

In addition to the above limitations, all magnetic field models are sensitive to the imperfectly known photospheric boundary condition.  Relatively small differences in the photospheric field, particularly in the flux balance, can lead to large changes in the resulting extrapolation \citep{riley14}.  Photospheric magnetograms can vary between observatories, due to differences both in the instrumentation and in processing techniques \citep{riley14}.  Synoptic magnetograms, on which global coronal models are usually based, are created from several magnetograph observations over the course of a Carrington rotation, typically neglecting surface flows.  Since observatories only see one side of the Sun at any given time, flux emerging on the far side of the Sun may not be incorporated into the synoptic magnetograms for several days.  When newly emerged flux \emph{is} incorporated into the synoptic magnetogram, the trailing polarity may not be included right away, leading to an apparent global flux imbalance which may be 'corrected' for in a variety of damaging ways.  Additionally, when the observatory is out of the ecliptic one pole of the Sun will not be observed at all during the Carrington rotation, and flux values in the polar region have to be approximated from earlier data or assumed to be zero.  Flux transport models \citep{schrijver03,arge10,arge13,hickmann15} offer state-of-the-art methods for evolving the flux in unobserved regions, but still do not seem to find a 'ground truth' about the photospheric magnetic field (e.g. see Figures $3$ and $4$ of \citet{hickmann15}).  For these reasons, magnetograms must be used judiciously, and it is unfortunate that coronal models are dependent on them.

Likely due to the many issues discussed above, it often occurs that coronal magnetic field models do not hold up well in comparison with coronal images.  The model may include more or less open flux than the images suggest, or may include open flux in regions where closed flux is observed.  Coronal images can give us a great deal of physical insight into the morphology of the magnetic field in the corona.  Recent advances in processing of eclipse images reveal astonishingly detailed information about the coronal structure (see \citet{druckmuller14, habbal14}.)  However, images cannot be used to give a value for the magnetic flux at a particular location, or to determine the \ac{3d} magnetic topology to a sufficient degree of accuracy to connect \emph{in situ} observations to photospheric or even coronal sources; for that, we need a \ac{3d} model.  In this paper, we describe a technique we have developed to combine the information available from coronagraph images with coronal magnetic field models to exploit the benefits of each.

There have been previous studies that utilized coronal \ac{euv} images in combination with models to try to more accurately determine the magnetic field.  Conlon and Gallagher \citep{conlonandgallagher10} used \ac{euvi} images to choose the best value of the parameter $\alpha$ in an \ac{lfff} model of a coronal active region.  Aschwanden and co-authors \citep{aschwanden13, aschwanden13b, aschwanden10} created \ac{nlfff} models of active regions by forward-fitting the underlying magnetograms with sets of buried magnetic monopoles and selecting the corresponding $\alpha$ values for each monopole that produced the greatest agreement with coronal loops traced in \ac{euvi} images.  Malanushenko and co-authors \citep{malanushenko12,malanushenko09} have presented a method for deriving an \ac{nlfff} model from sparsely-distributed \ac{euv} loop observations.  These studies have focused on modelling the field in active regions, where the high density results in bright \ac{euv} emission and the limited size of the region allows for a reasonable number of free parameters.  Additionally, the complexity of these methods suggests that their primary application would be the detailed study of a specific region of interest, rather than casual production of models for programmatic use.  In contrast, the purpose of our study has been to develop a method for the fast production of global coronal magnetic field models based on widely available synoptic magnetograms and coronal images.  Global models are necessary for studies on a wide array of topics, including the connectivity of active regions \citep{schrijver13,tadesse12}, the topology of the corona through the solar cycle \citep{platten14,wang03}, and as a larger context for studies of coronal activity in localized regions \citep{schrijver13,conlonandgallagher10}.  They are also instrumental for global heliospheric simulation and interpreting \emph{in situ} measurements by upcoming near-Sun missions \ac{so} and \ac{spp}.

The remainder of the paper is laid out as follows.  In section \ref{methods} we describe two approaches we have used for optimizing the agreement between a \ac{pfss} model and image-based constraints.  Section \ref{testing} describes some theoretical tests to which we have applied our method.  In section \ref{discussion} we discuss what we have learned about the limitations and applicability of our method, and section \ref{conclusions} summarizes our conclusions.
\section{Methods}
\label{methods}
The numerical algorithms presented below can, in principle, be applied to any of the field models described above: \ac{lfff}, \ac{nlfff}, or even \ac{mhd} models.  All that is required is to be able to calculate the field components on a \ac{3d} grid for comparison with the constraints.  The primary obstacle to using the method with \ac{mhd} models is the computational cost - because the optimizing algorithm is searching the solution space directly, it needs to calculate the field for a large number of magnetogram configurations.  

In the interest of creating quick, global models, we chose to apply our optimization method to a \ac{pfss} model, leaving application to more sophisticated models for future studies.  Optimizing \ac{pfss} models is a convenient starting point since the outcomes are fully defined by the photospheric boundary condition which we intend to vary, leaving us with no intrinsic free parameters.  The optimization begins with a synoptic magnetogram and a set of angles $\{\theta_o\}$ that describe the orientation of magnetic features at locations $\{\boldsymbol{r}\}$.  (In real-world applications these angles would best be derived from coronagraph images, but for this study we have used artificial constraints based on theoretical test problems.)  The \ac{pfss} extrapolation is calculated using the \ac{pfss} library from SolarSoft; this results in three data arrays giving the three components of the magnetic field on a 3D spherical grid.  The optimization software then samples the \ac{pos} field at the specified $\{\vec{r}\}$, giving a set of model orientation angles $\{\theta_m\}$. It then calculates the penalty function $J$ as follows:
\begin{equation}
\label{penalty}
J=\beta \sum_{k=1}^N (\theta_{o,k} - \theta_{m,k})^2+\gamma\,  O
\end{equation}
where $\beta$ is a normalization constant meant to keep $J$ from becoming excessively large when many constraints are being used, $\theta_{o,k}$ and $\theta_{m,k}$ are the $k^{th}$ observed and model field orientations, and $O$ represents optional terms that incorporate additional knowledge about the solution.  Example terms $O$ might include a regularization term \citep{bickel06}, to encourage solutions with smooth magnetograms, or a net flux term, to encourage solutions that have little or no net flux; in these cases the constant $\gamma$ must be chosen so as to balance the requirement to match the original constraints against these additional objectives.  $O$ may even incorporate constraints that reflect additional knowledge about the field, such as flux measurements garnered in a particular region.  $J$ decreases as the model $\{\theta_m\}$ approach the observed values; the lower $J$ is, the better the agreement between the model and the image-based constraints.  

We use the well-known downhill simplex method \citep{nelder65} to minimize $J$, finding the boundary condition (magnetogram) that gives optimal agreement between the model field and the coronagraph images.  The algorithm starts by defining a simplex - a set of $m+1$ vertices in the $m$-dimensional solution space ($m$ here is the number of pixels in the magnetogram being optimized).  Each vertex is a different realization of the magnetogram (typically each starting vertex is equivalent to the original magnetogram perturbed in one pixel), and the software calculates an extrapolation and a resulting penalty function for each vertex.  It then iteratively perturbs the worst point in the simplex according to a set of simple rules, until it eventually finds a minimum and contracts to a point; the optimization terminates when the size of the simplex falls below a user-defined value.  For reasons discussed below, we have also allowed the user to terminate the program after a specified time has elapsed.  

The downhill simplex method is an example of a direct search optimization algorithm \citep{kolda03}.  Direct search methods have the advantage that they do not require the computation or estimation of the derivatives of the penalty function (difficult in our case), but they are vulnerable to finding local (rather than global) minima.  In general, direct search methods become less efficient and are increasingly vulnerable to finding local minima as the dimensionality of the solution space is increased.  Since each pixel in the magnetogram is an additional dimension in the solution space, each increase in the angular resolution of input magnetograms greatly increases the computation time and may decrease the quality of the result.  This inspired us to develop a second, similar version of our code that uses the same basic optimization scheme but optimizes the magnetogram flux values in the \ac{sht} space.  While optimizing a relatively small number of the transform coefficients, we are able to make global changes to the magnetogram.  The \ac{pfss} software calculates the \ac{sht} of the magnetogram as a matter of course before extrapolating the field, so it is ideally suited for this approach.

The optimization in the transform space proceeds as follows.  Rather than forming a simplex from perturbed realizations of the magnetogram, we calculate the spherical harmonic transform of the magnetogram using the SolarSoft \ac{pfss} software, then form the simplex from a user-specified number $M$ of the lowest-order coefficients.  Initially, at each vertex in the simplex, one coefficient is perturbed.  The optimization software is able to deconstruct the simplex vertices as needed to reform a perturbed transform and the perturbed transform is incorporated into the appropriate common block variables used by the \ac{pfss} software.  At this point the model calculation can be performed, the penatly function calculated, and simplex vertices evolved as described above.  While we may choose to optimize only the first $N$ coefficients, all of the coefficients are retained and used in the field calculation.  This allows us to create models with angular resolution matching that of the original magnetogram even though we are optimizing at a lower resolution.

\section{Model Problems}
\label{testing}
Here we describe the performance of our technique on two model problems, one for the original version of the optimization code, which optimizes the magnetogram flux values, and one which was developed for testing the optimization of the \ac{sht}.
\subsection{Dipole Perturbation}
\begin{figure}
  \begin{tabular}{c c}
  \subfloat{
      \includegraphics[width=0.4\textwidth]{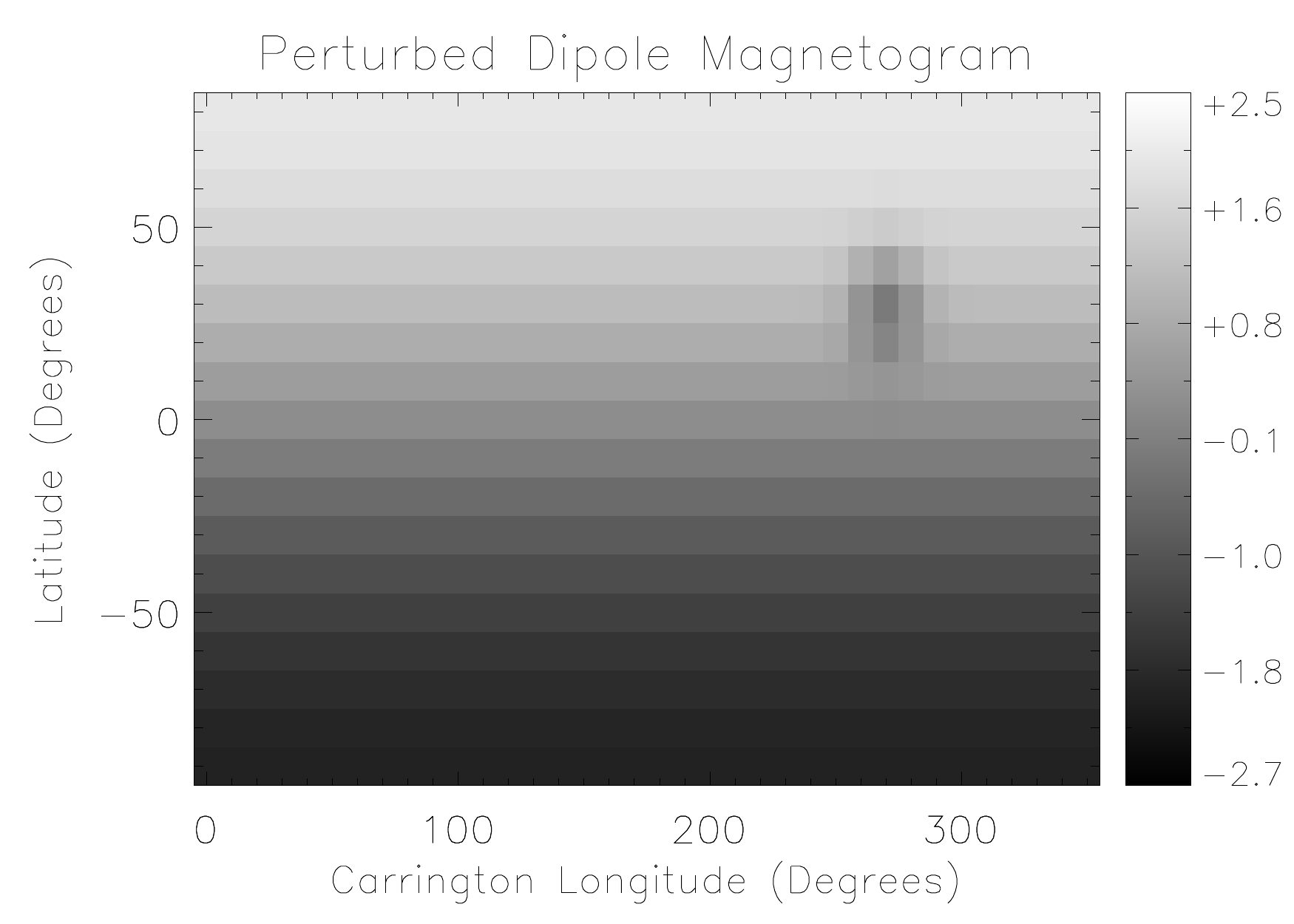}
      \label{dipole_input_magnetogram}} &
  \subfloat{
      \includegraphics[width=0.4\textwidth]{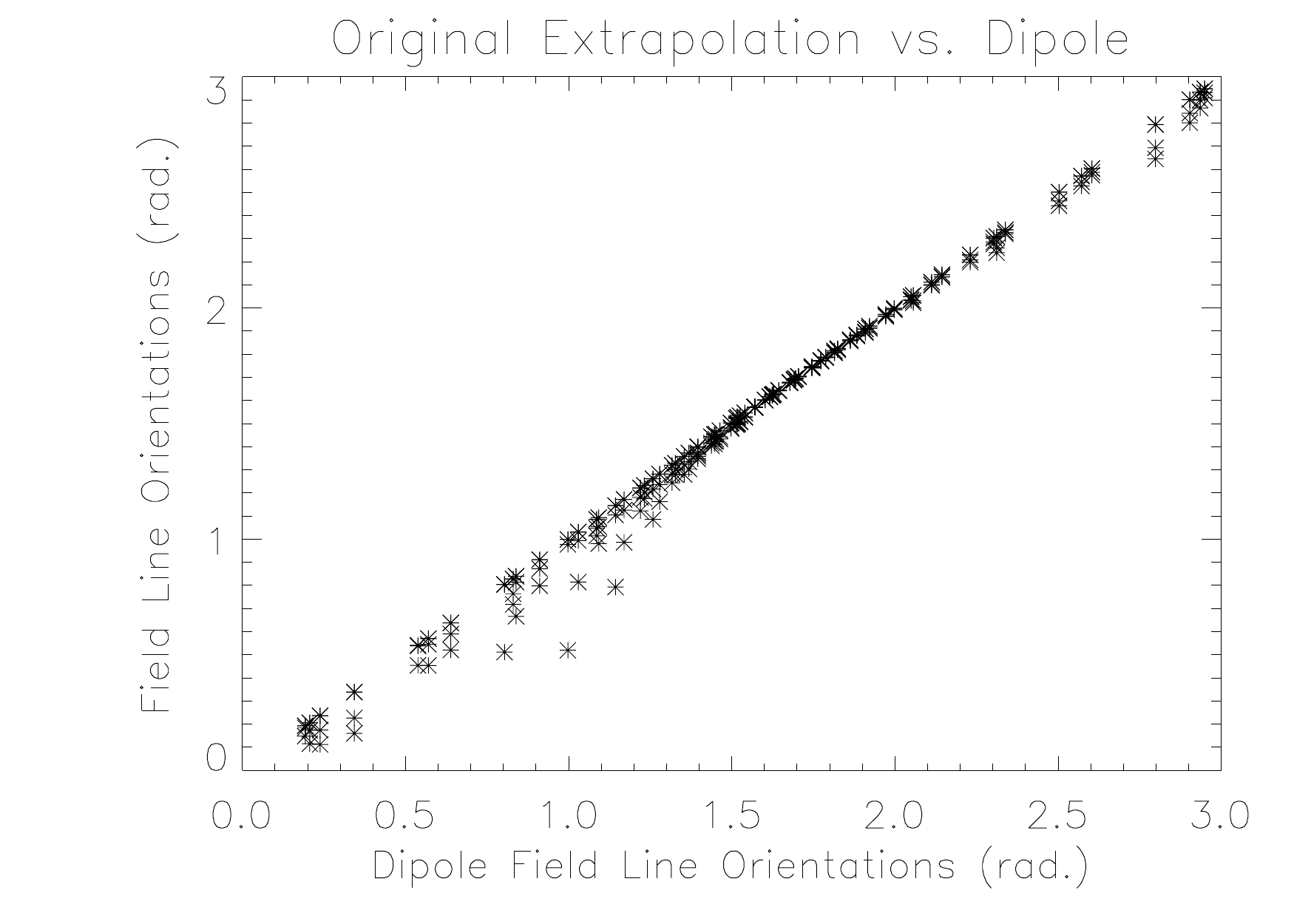}
      \label{dipole_input_components}}\\
  \subfloat{
      \includegraphics[width=0.4\textwidth]{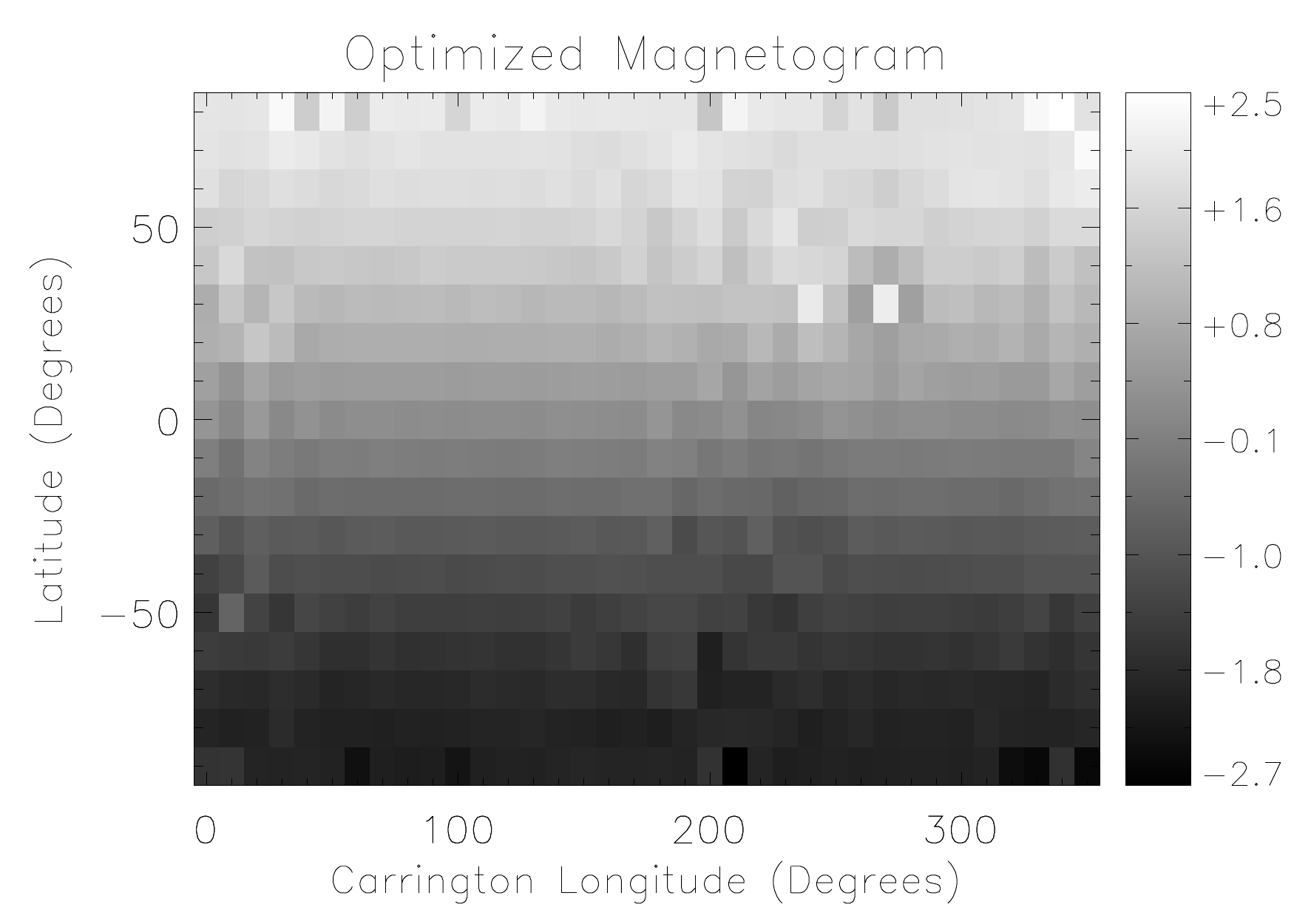}
      \label{dipole_output_magnetogram}} &
  \subfloat{
       \includegraphics[width=0.4\textwidth]{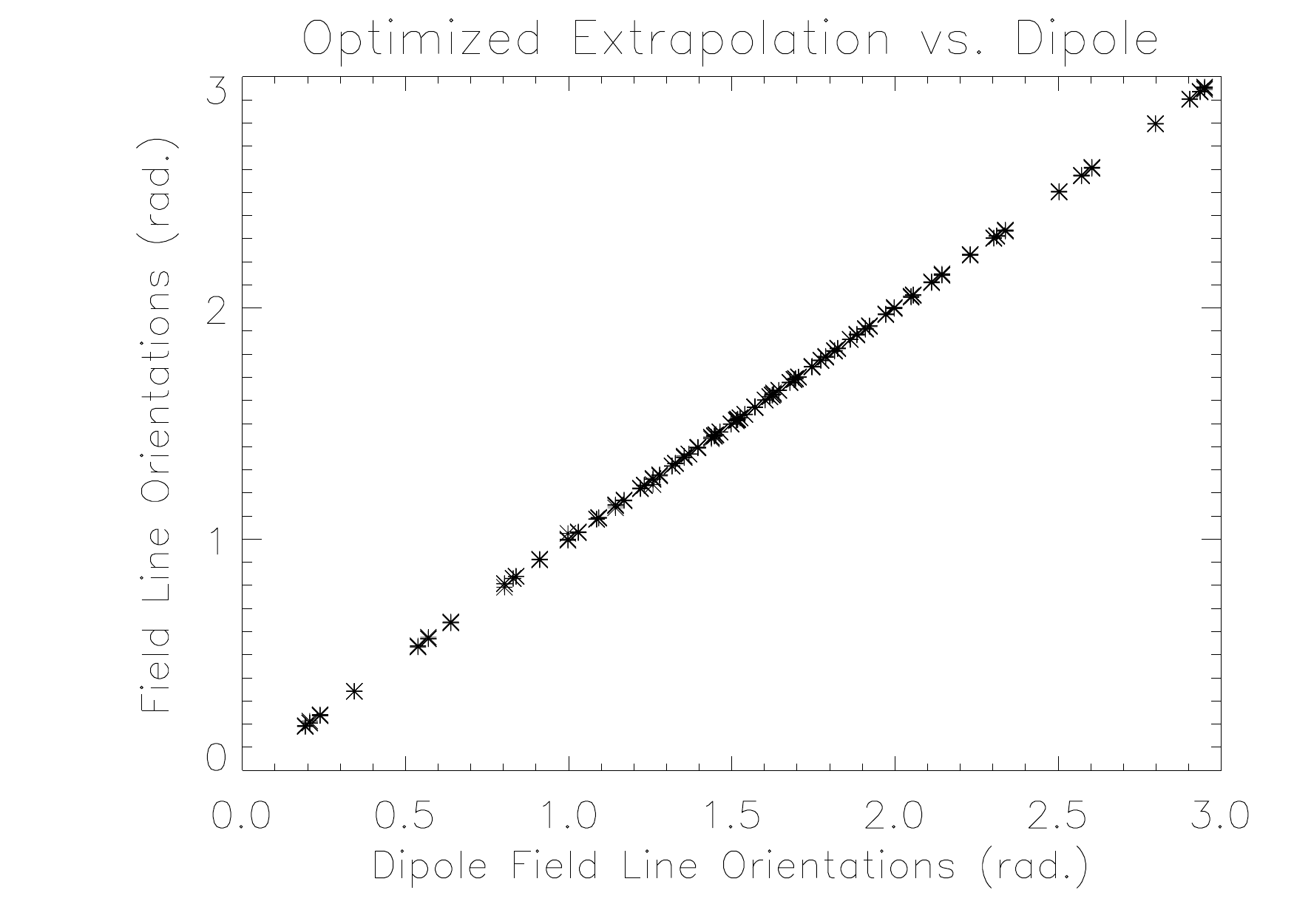}
       \label{dipole_output_components}}\\
  \subfloat{
      \includegraphics[width=0.5\textwidth]{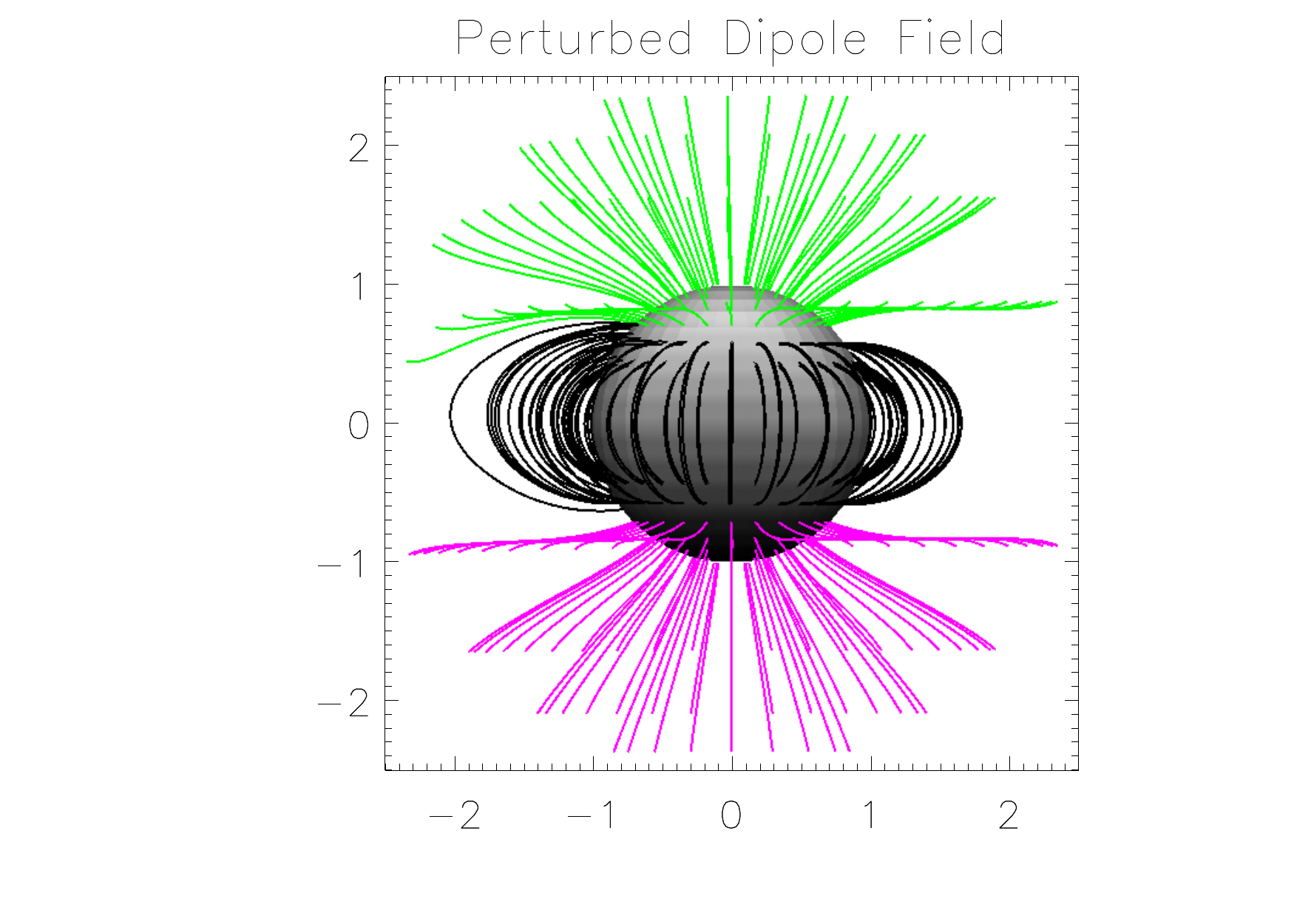}
      \label{dipole_input_hairyball}}  &
  \subfloat{
      \includegraphics[width=0.5\textwidth]{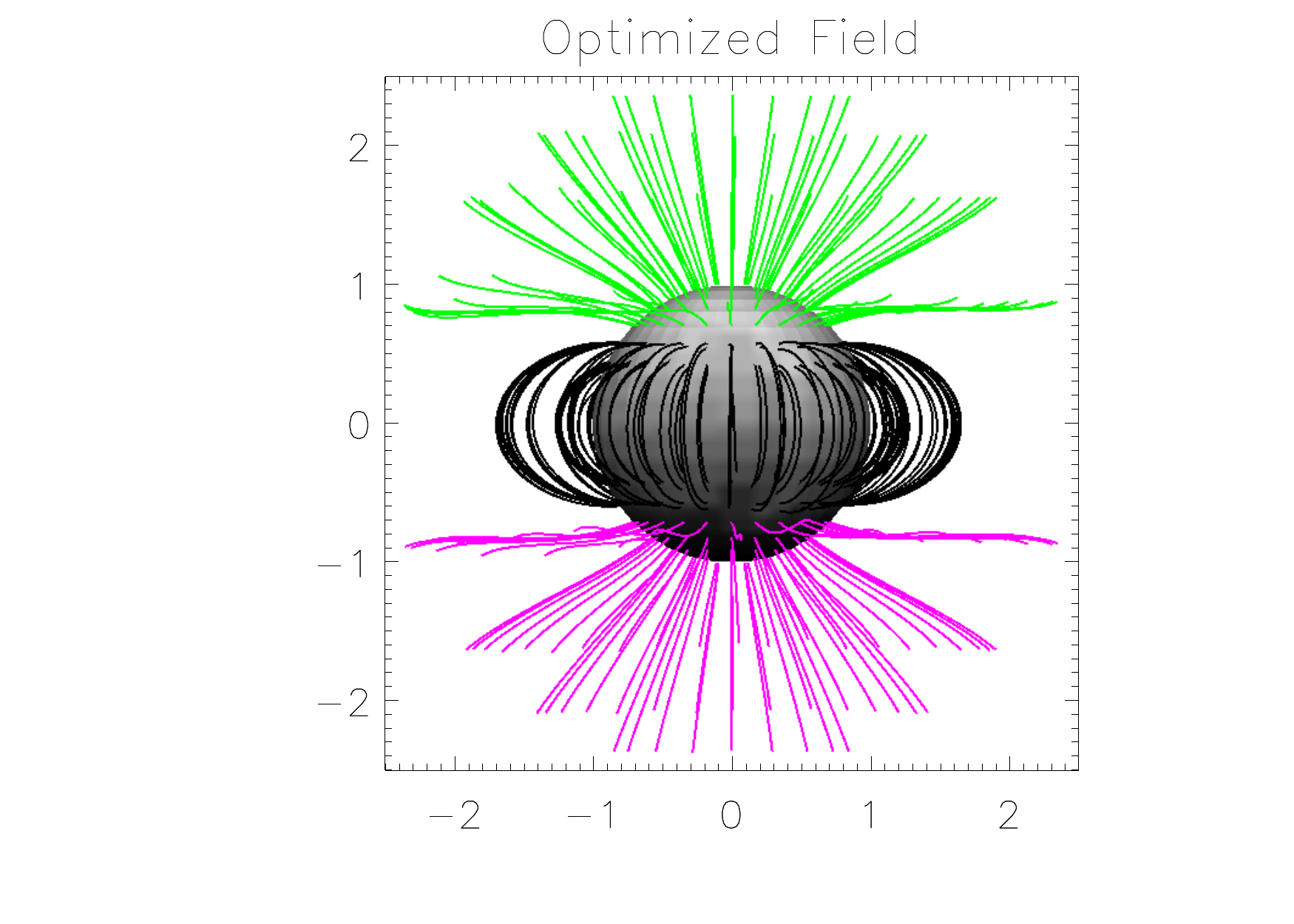}
      \label{dipole_output_hairyball}} \\
\end{tabular}
\caption{Removing a monopolar perturbation added to a dipolar magnetogram with the optimization method.  The perturbed magnetogram was input to our optimization software as an initial guess, along with purely dipolar constraints.  The top row shows the input magnetogram and a comparison of the field orientations $\{\theta_o\}$ in two image planes versus the dipolar constraints.  The middle row shows the same figures after optimization, where we can see much of the perturbation has been removed/counter-balanced and the field orientations now agree much better with the expected values.  In the bottom row we show a projected view of the extrapolated field before (left) and after (right) optimization, where the field has become visibly more symmetric after the optimization.}
\label{dipole_test}
\end{figure}
One subject of particular interest related to synoptic magnetograms is the decay and/or emergence of magnetic flux on the far side of the Sun.  This is troubling from an optimization standpoint because it means that the optimal magnetogram may diverge dramatically from the initial suggestion we have provided to the software.  To test the performance of the optimization software on problems of this type, we created an artificial dipolar magnetogram and added to it a negative monopolar perturbation.  The perturbation magnitude is Gaussian, with a maximum amplitude that is a little over $33\%$ of the dynamic range of the magnetogram, and is truncated outside of a square of length 5 pixels centered at $270^{\circ}$ longitude, $34^{\circ}$ latitude.  We provided this perturbed dipolar magnetogram to the optimization software, but with constraints $\{\theta_o\}$ that corresponded to an ideal dipole field configuration.  Figure \ref{dipole_input_magnetogram} shows the input perturbed magnetogram (treated by the software as an initial guess) and \ref{dipole_input_hairyball} shows the resulting \ac{3d} field extrapolation.  The effect the perturbation has on the field lines can be seen by comparing the left and right hemispheres of Figure \ref{dipole_input_hairyball} - the open field lines on the left near the north pole are bent downward, and the closed field lines are more pinched near the equator.  In Figure \ref{dipole_input_components} we have plotted the model field orientations $\{\theta_m\}$ from every latitudinal and radial bin in two azimuthal planes, versus the $\{\theta_o\}$ expected for a dipolar field configuration.  If the extrapolation produced a dipole field, the points in Figure \ref{dipole_input_components} would form a straight line of slope one.

Optimization based on this perturbed magnetogram took approximately thirty minutes on an older desktop computer (Intel Core 2 6700 processor at 2.66 GHz).  Figure \ref{dipole_output_magnetogram} shows the magnetogram after optimization and \ref{dipole_output_hairyball} shows the model field based on this optimized magnetogram.  It can be seen that the field is now much more symmetric from left to right, and the perturbation has been greatly reduced by the optimization.  

\begin{figure}
  \begin{tabular}{ c c }
  \subfloat{
       \includegraphics[width=0.45\linewidth]{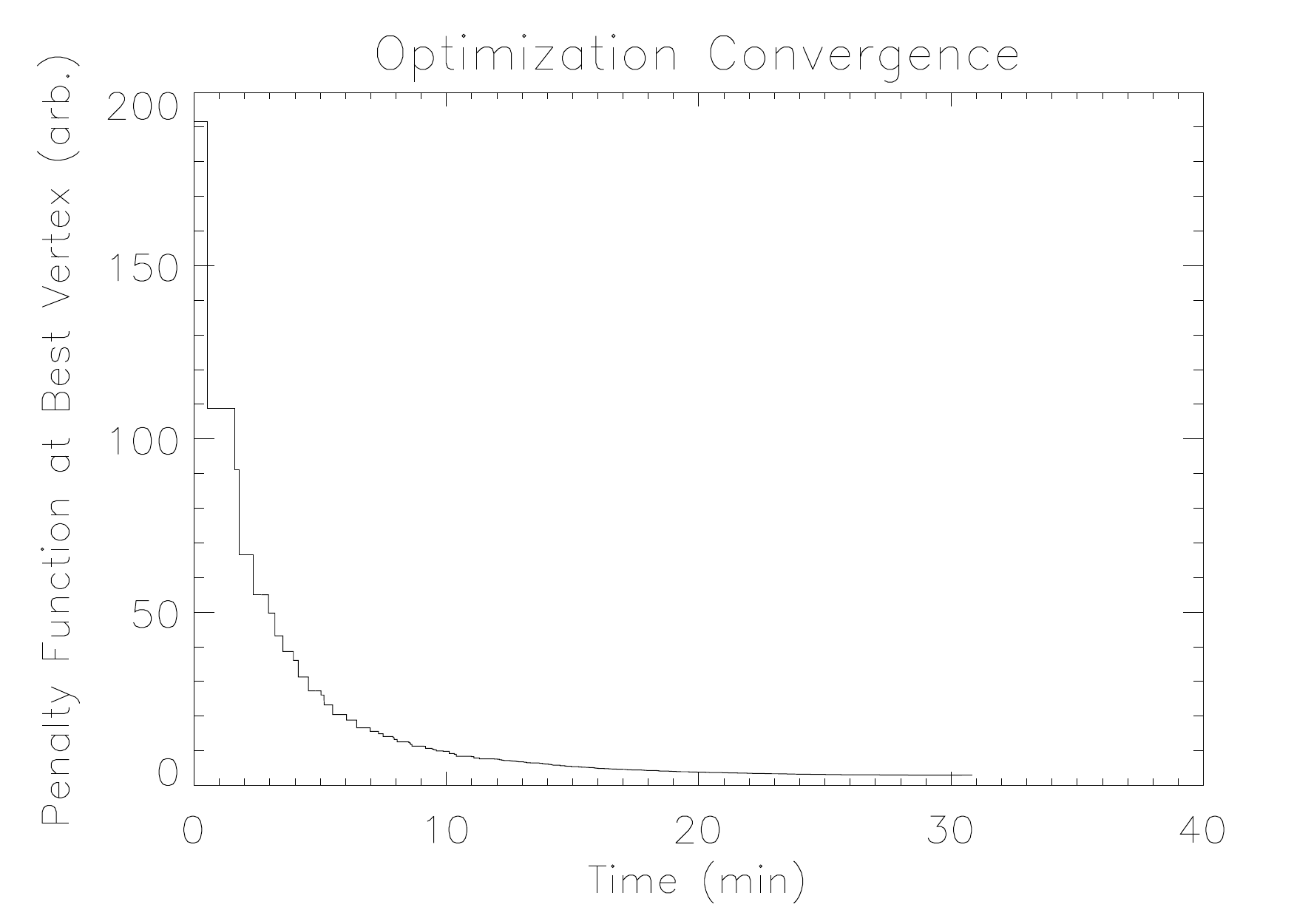}
      \label{dipole_convergence}} &
  \subfloat{
      \includegraphics[width=0.45\linewidth]{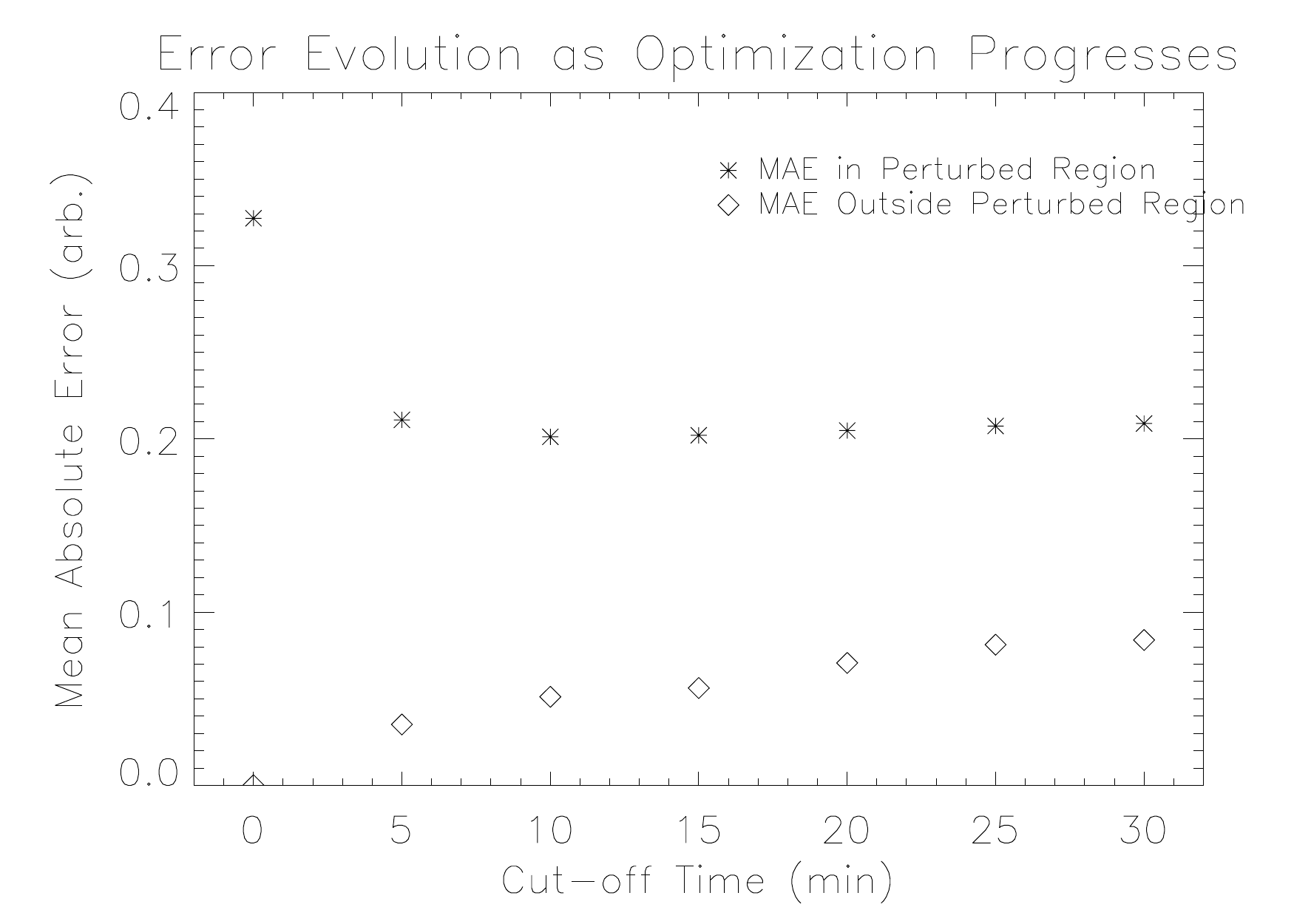}
      \label{dipole_mae}}\\
 \end{tabular}
\caption{(Left) The value of $J$ for the perturbed dipole test as a function of computation time.  (Right) The \ac{mae} between the ideal dipole magnetogram and the best vertex in the simplex, separated into a component in the region of the perturbation (asterisks) and everywhere else (diamonds).  During the early stages of the optimization, $J$ improves rapidly as the \ac{mae} in the perturbation region drops, then slowly declines while the \ac{mae} in the remainder of the magnetogram rises steadily.}
\end{figure}

There is some noise added throughout the magnetogram after the optimization.  This happens because introducing a small amount of random noise to a dipolar magnetogram doesn't produce much change in the $\{\theta_m\}$ - the model field is still roughly dipolar.  Figure \ref{dipole_convergence} shows the value of the penalty function at the best point in the simplex as a function of computation time.  The long, slow tapering seen in the latter stages of the optimization is where most of this noise is added; as the optimization proceeds the magnetogram drifts further from the initial guess.  Figure \ref{dipole_mae} illustrates this; here we have plotted the \acf{mae} in the region where the perturbation is and in the remainder of the magentogram as a function of computation time.  The \ac{mae} in the perturbed region falls rapidly at first, then much more gradually, while the \ac{mae} elsewhere in the magnetogram climbs steadily for most of the optimization time.  We found that adding additional image planes reduces the spatial noise caused by the optimization without significantly increasing the computation time.  We also incorporated an option into the software to allow for termination after a user-specified period of time.  However, we note that the growth of the noise is limited by the fact that it must not create a significant increase in the penalty function (\emph{i.e.} it must not significantly affect the resultant model field.)
\subsection{Harmonic Perturbation}
 \begin{figure}
 \begin{tabular}{ c c }
   \subfloat{
        \includegraphics[width=0.45\textwidth]{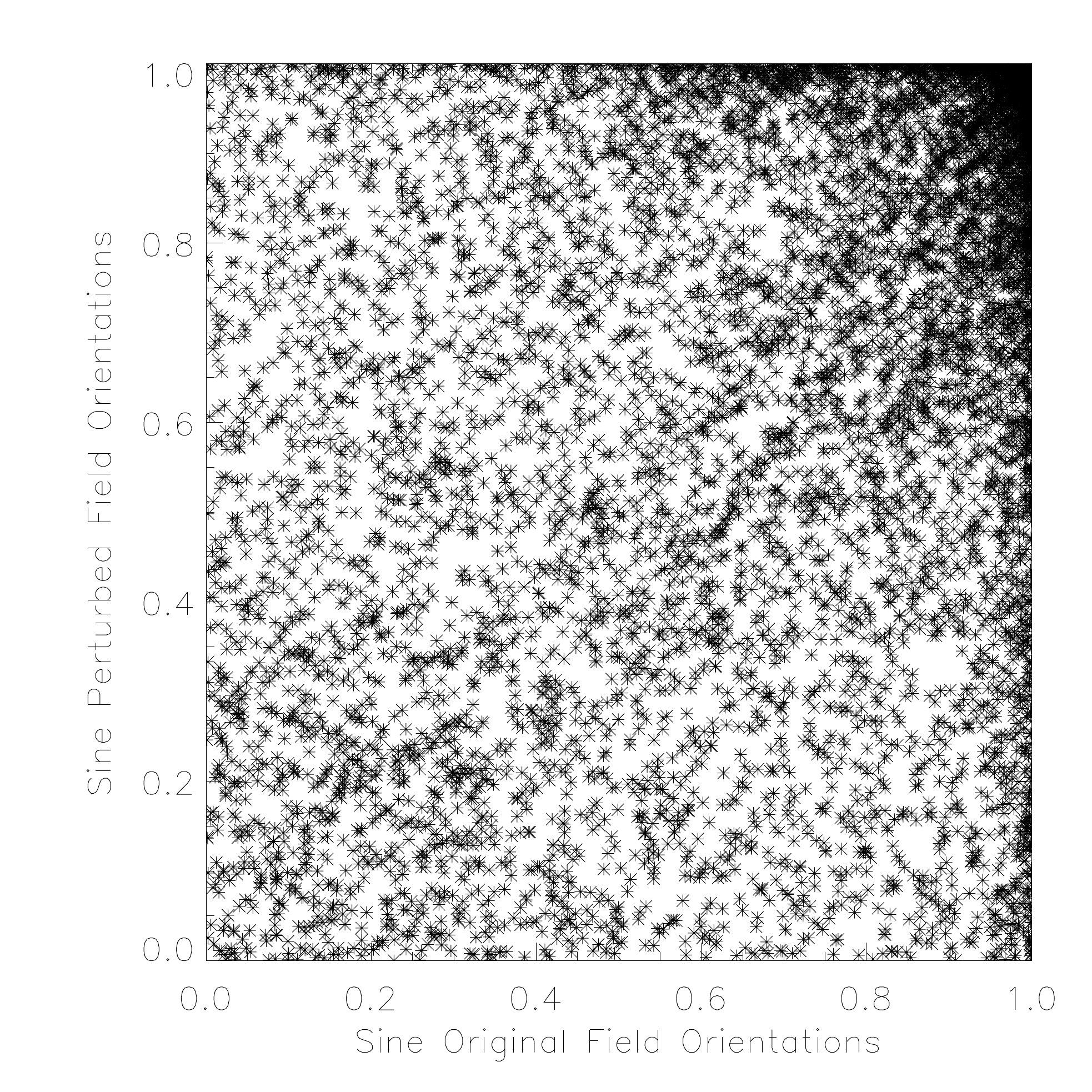}
        \label{harmonic_components_orig}} &
   \subfloat{ 
       \includegraphics[width=0.45\textwidth]{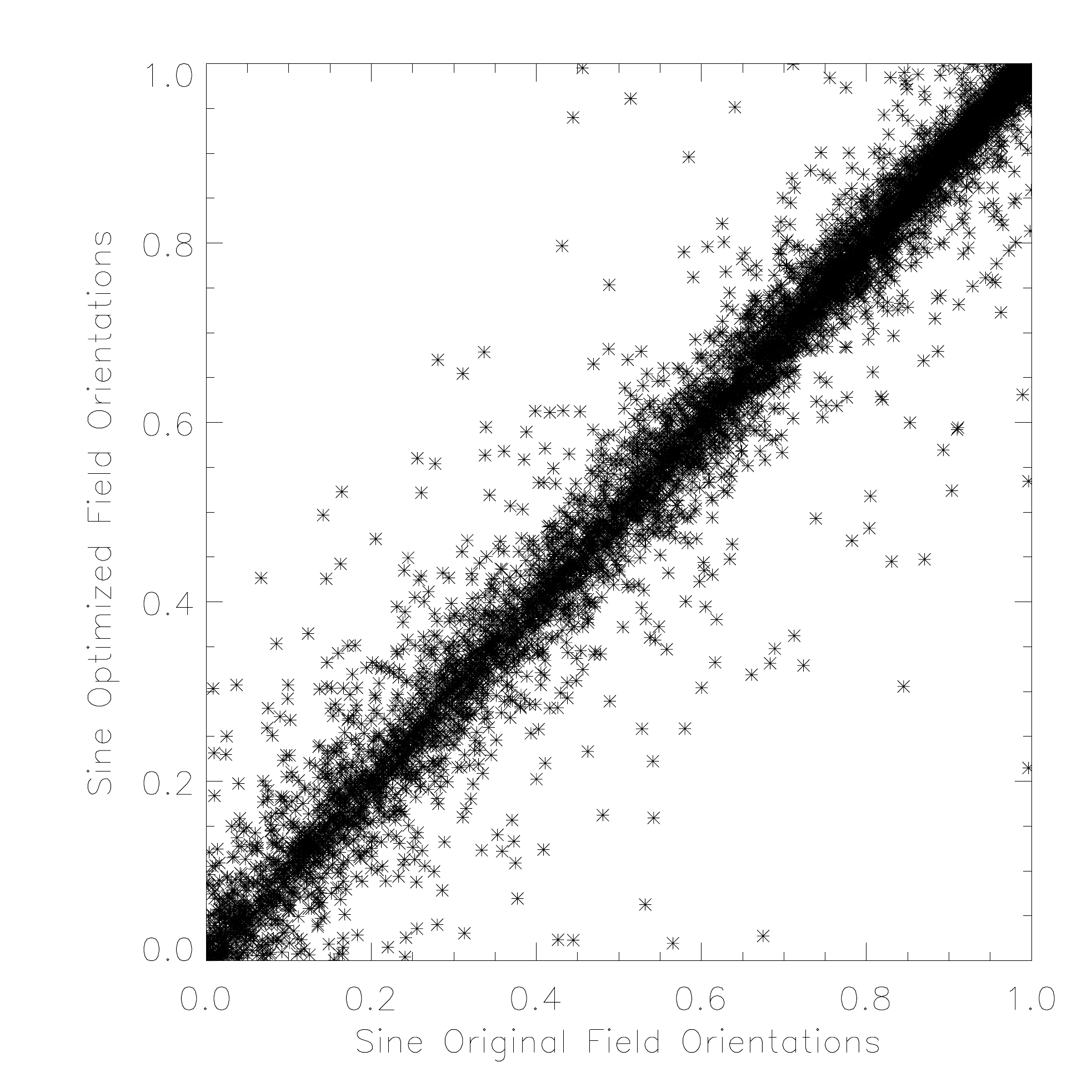}
       \label{harmonic_components_opt}}\\
 \end{tabular}
 \caption{Magnetic field orientation angles in the perturbed (left) and optimized (right) magnetic field solutions versus the angles in the unperturbed model.  Sines are plotted due to the periodic nature of the angles.  The agreement in the optimized solution is clearly much closer to the ideal.}
 \label{harmonic_components}
\end{figure} 
\begin{figure}
 \label{harmonic_test}
  \begin{tabular}{ c c c }
  \subfloat{
       \includegraphics[width=0.3\textwidth]{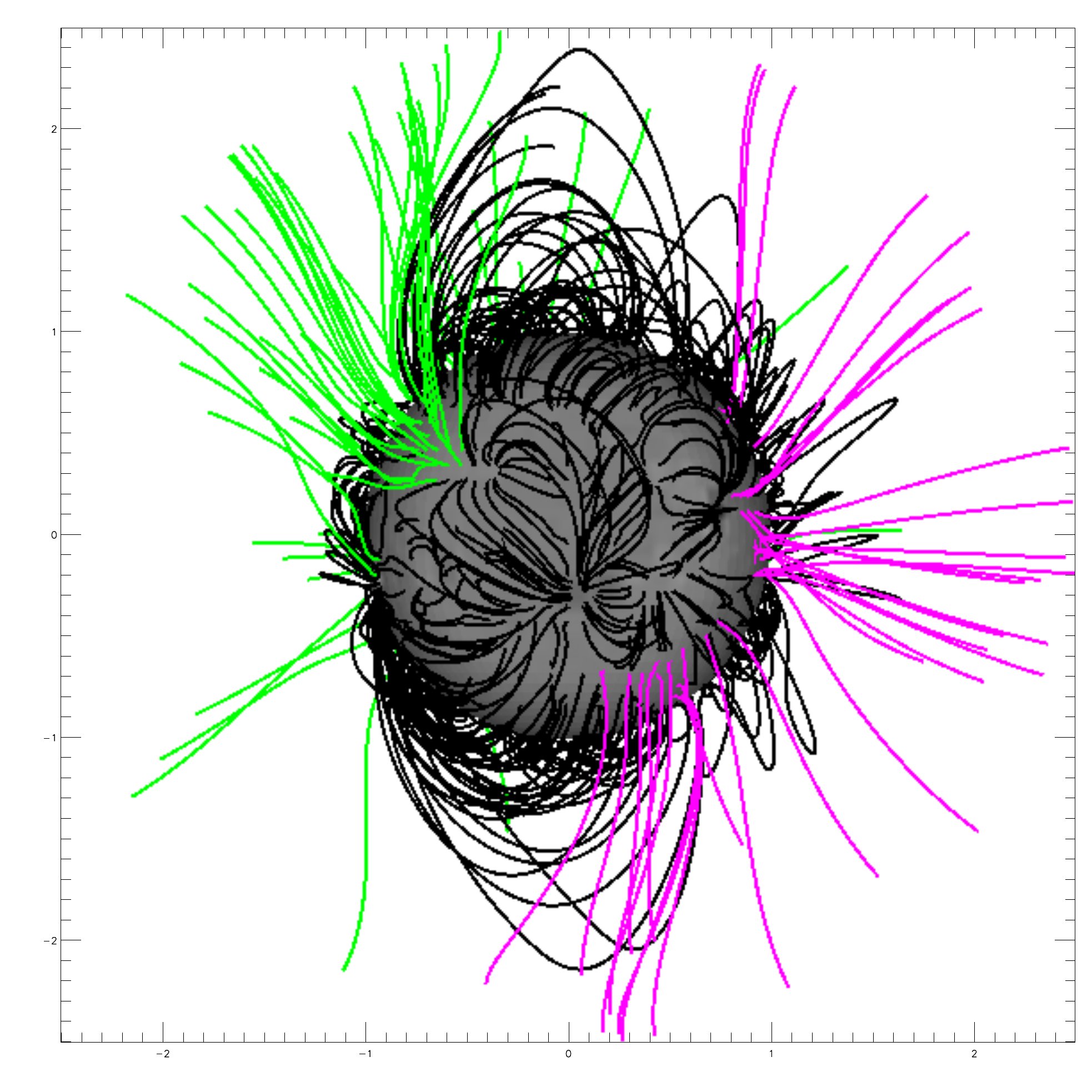}
       \label{harmonic_hairyball_omag_0}} &
  \subfloat{
      \includegraphics[width=0.3\textwidth]{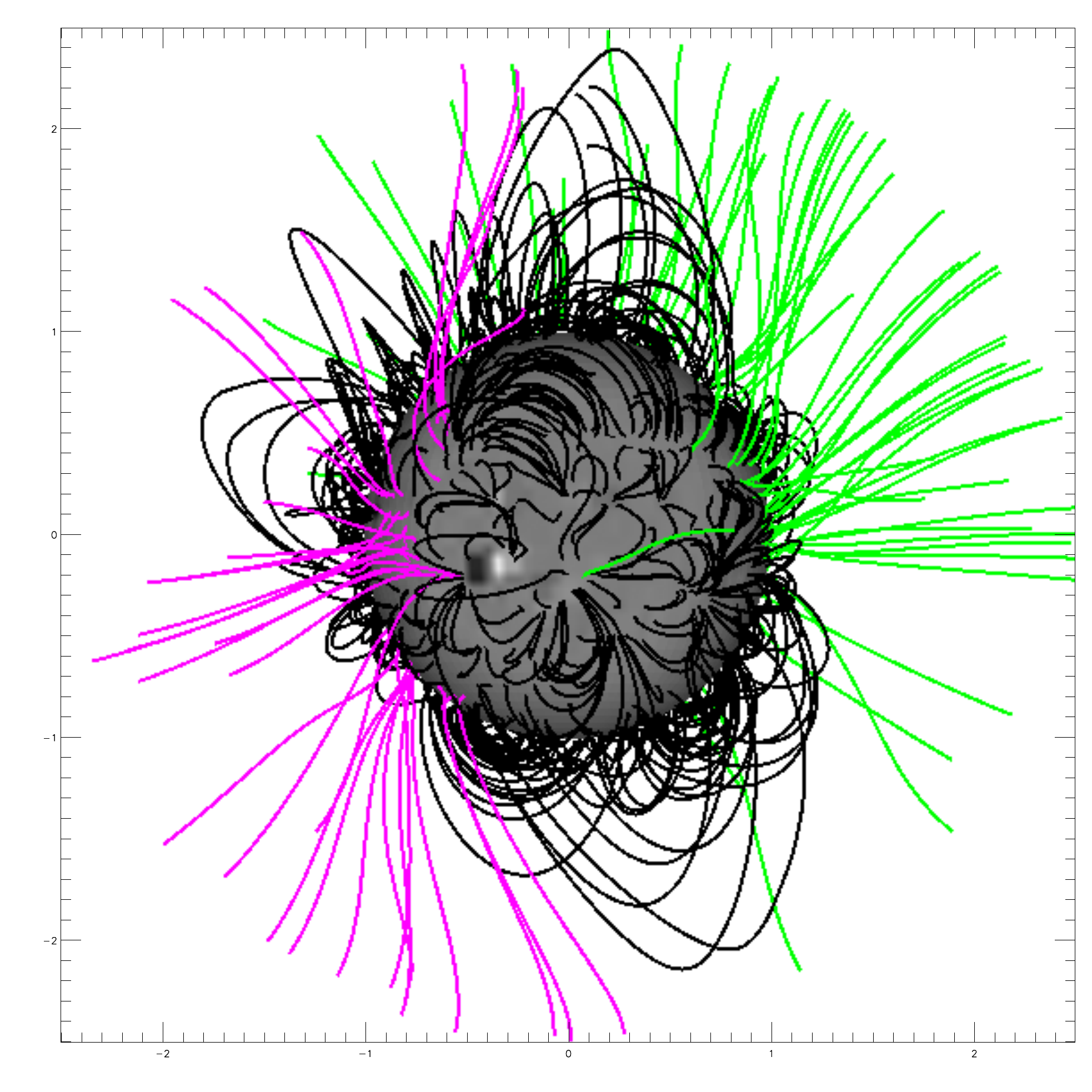}
      \label{harmonic_hairyball_omag_120}} &
  \subfloat{
      \includegraphics[width=0.3\textwidth]{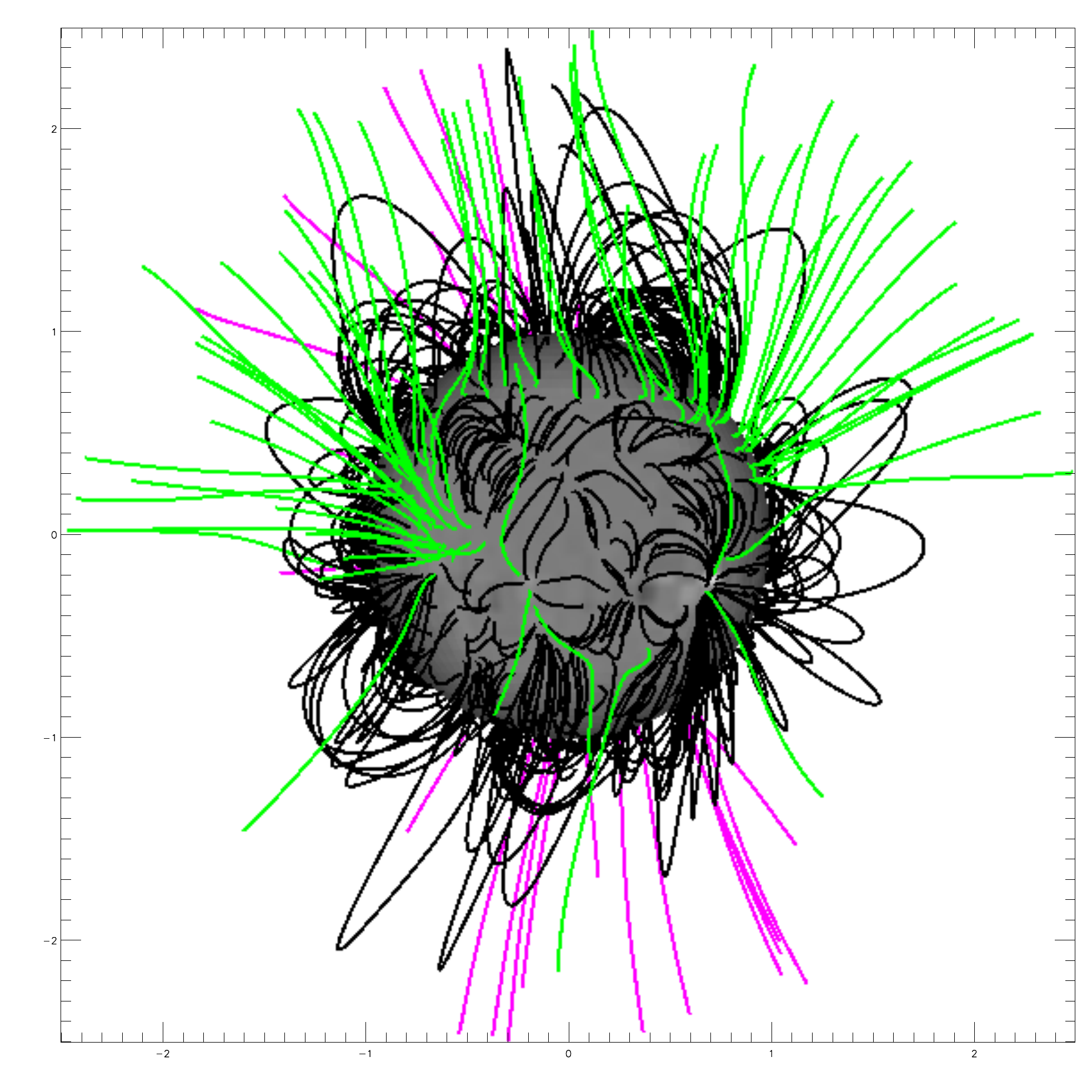}
      \label{harmonic_hairyball_omag_240}}\\
  \subfloat{
      \includegraphics[width=0.3\textwidth]{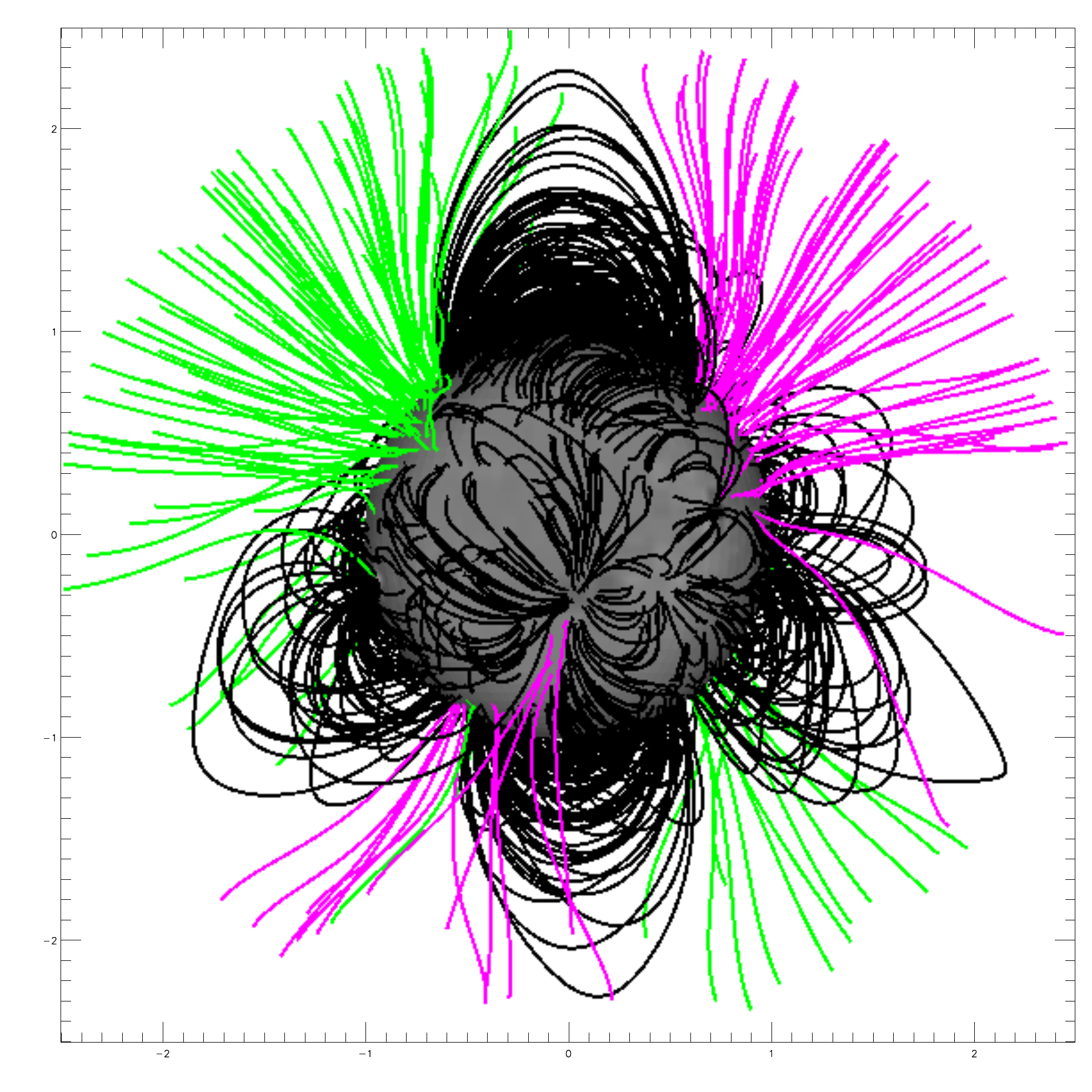}
      \label{harmonic_hairyball_pert_0}} &
  \subfloat{
      \includegraphics[width=0.3\textwidth]{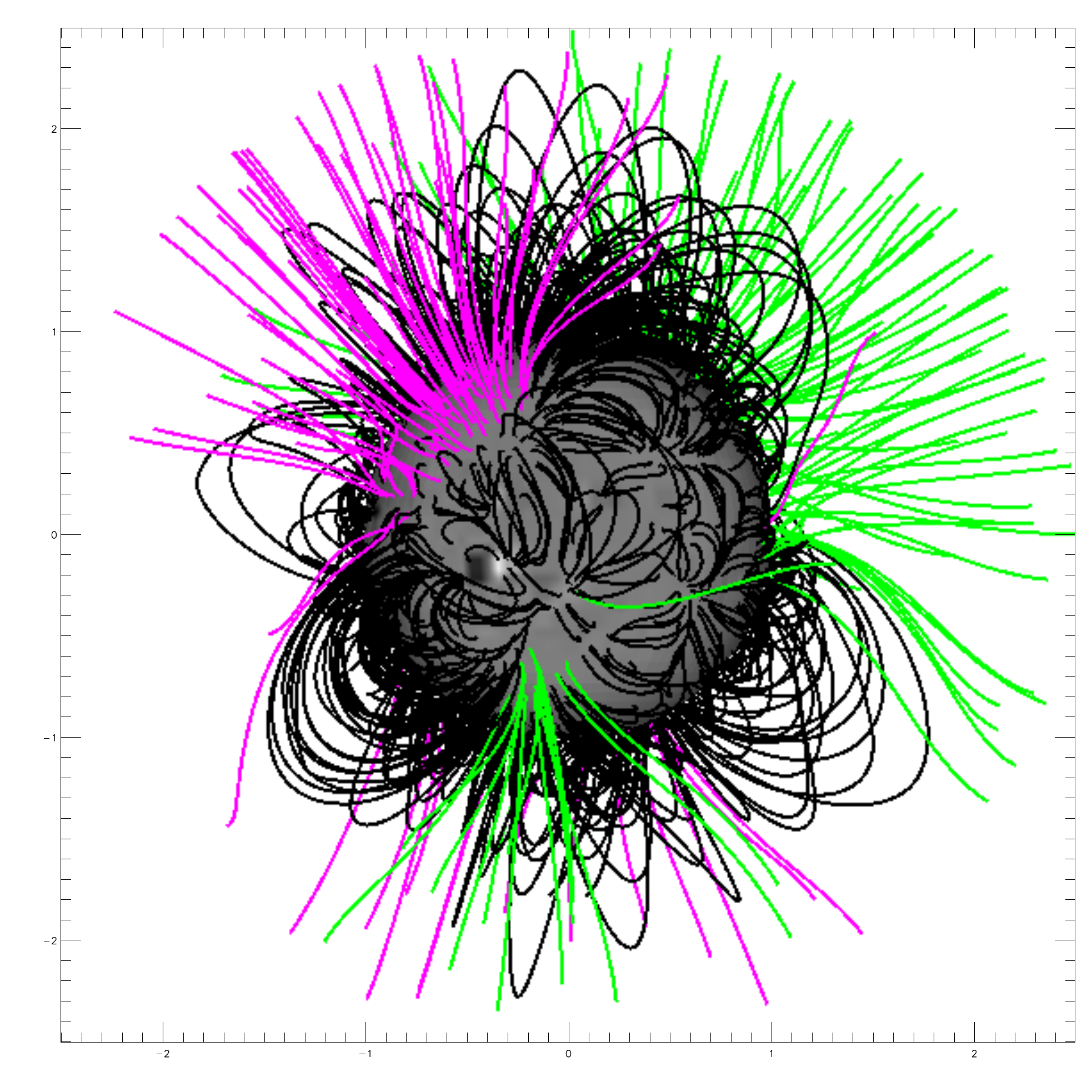}
      \label{harmonic_hairyball_pert_120}} &
  \subfloat{
      \includegraphics[width=0.3\textwidth]{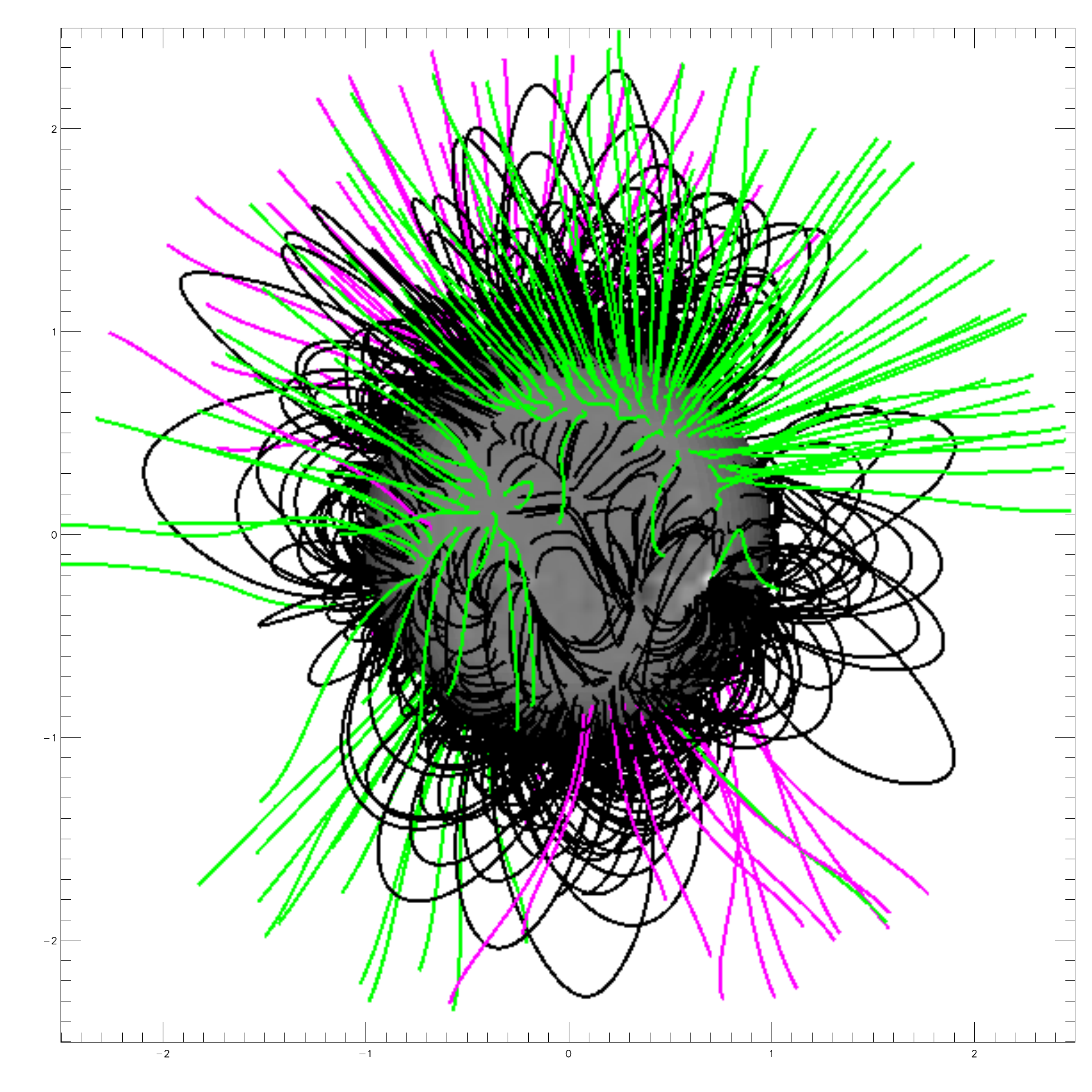}
      \label{harmonic_hairyball_pert_240}}\\
  \subfloat{
      \includegraphics[width=0.3\textwidth]{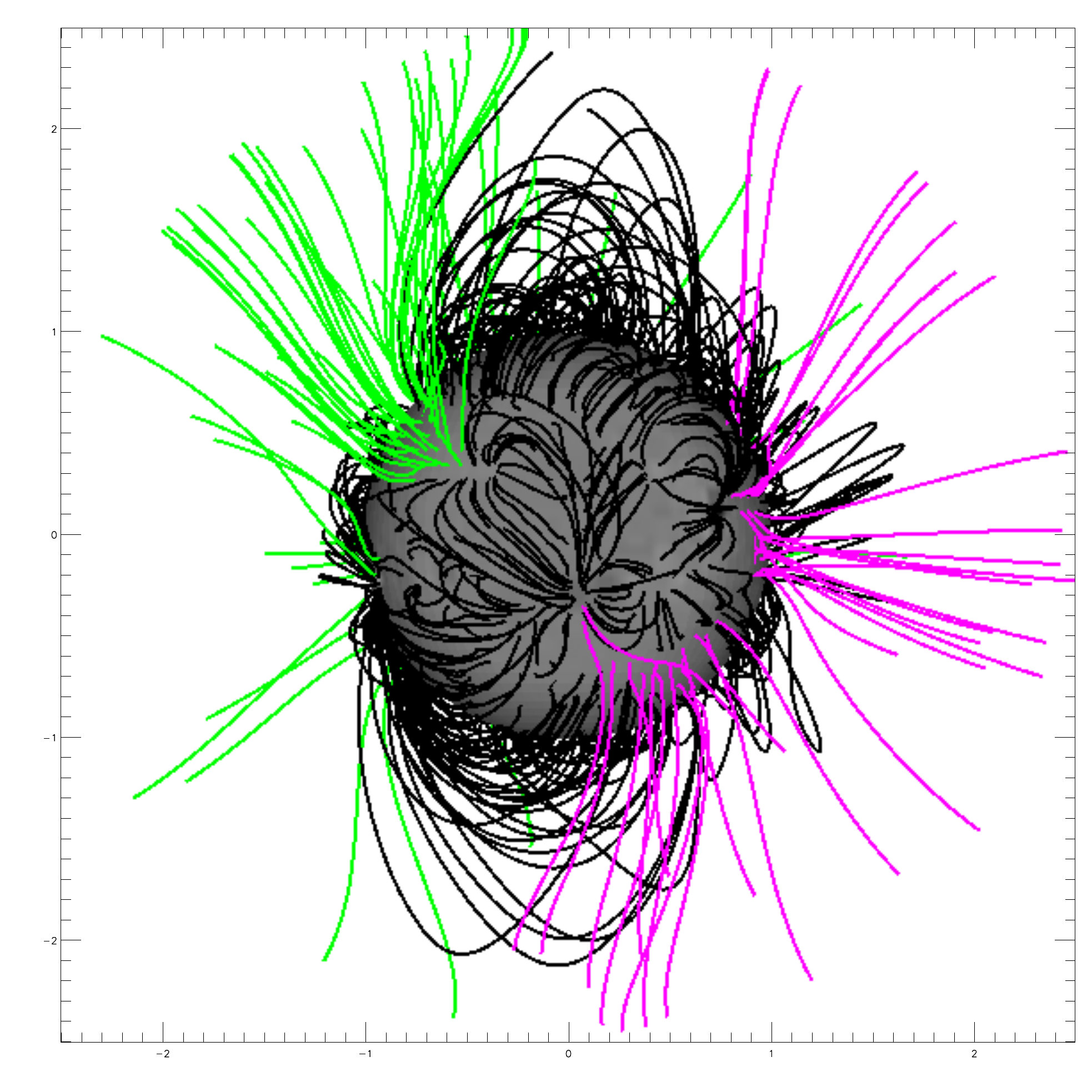}
      \label{harmonic_hairyball_opt_0}} &
  \subfloat{
      \includegraphics[width=0.3\textwidth]{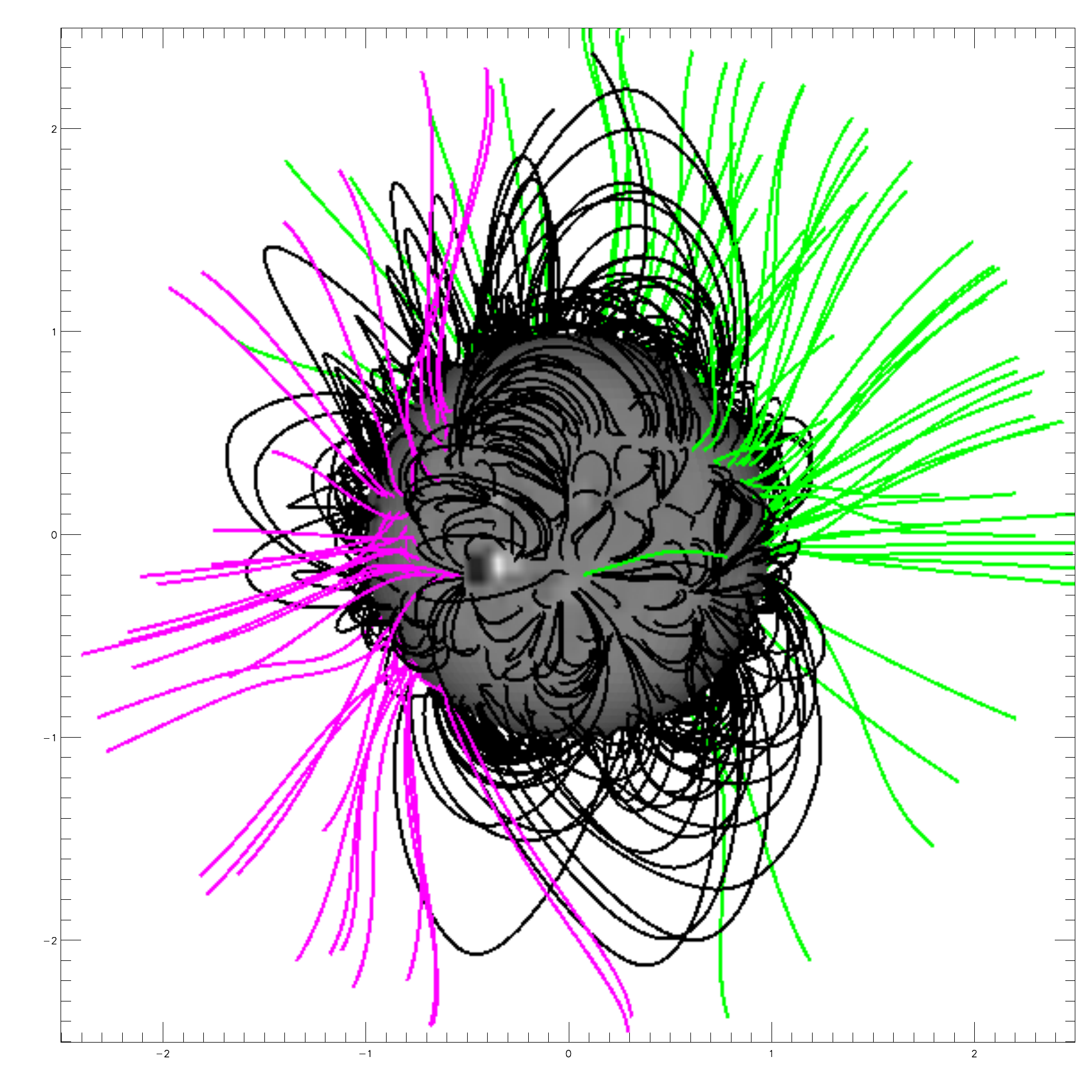}
      \label{harmonic_hairyball_opt_120}} &
  \subfloat{
      \includegraphics[width=0.3\textwidth]{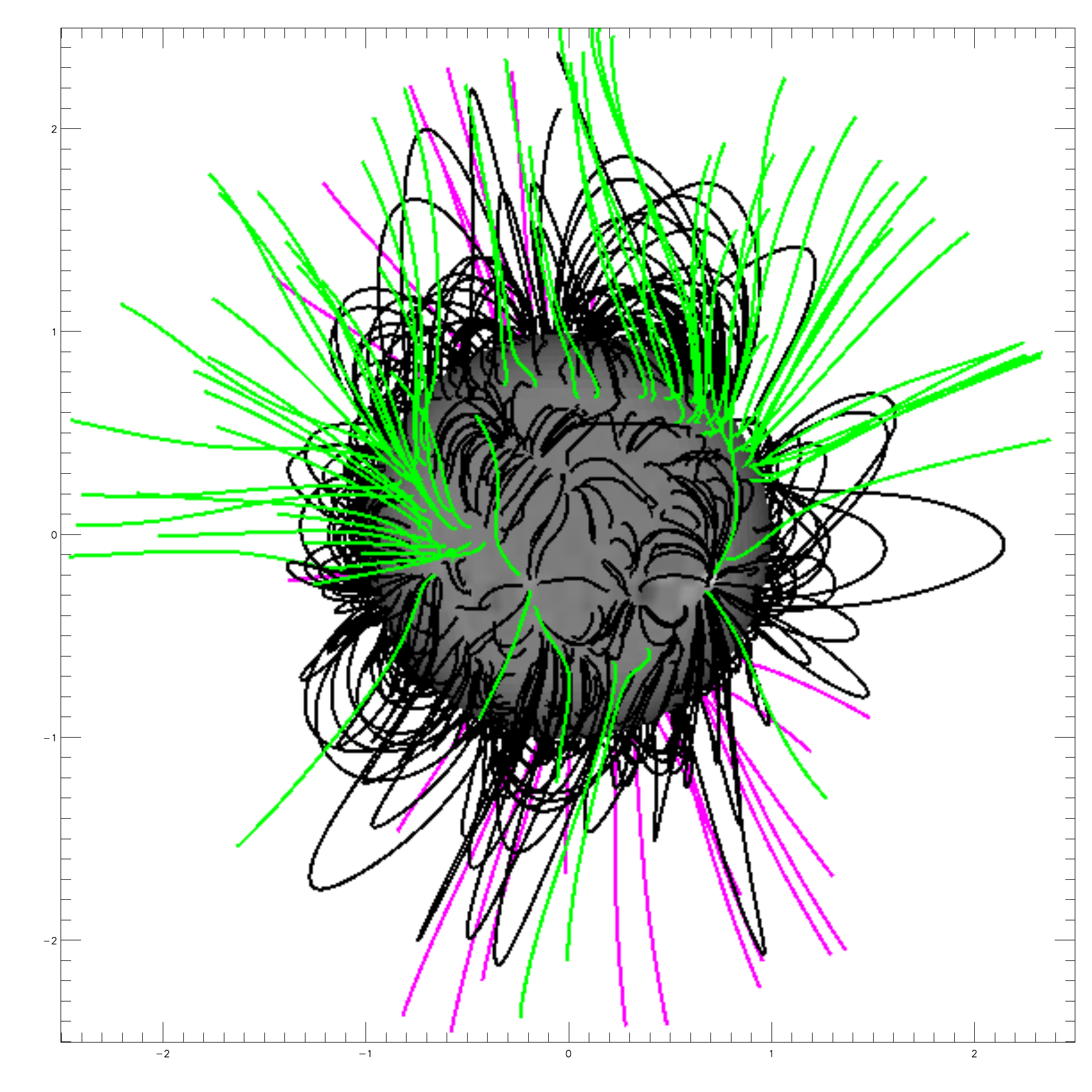}
      \label{harmonic_hairyball_opt_240}}\\
 \end{tabular}
\caption{Images showing the unperturbed \ac{pfss} model magnetic field (top row), the field after perturbation of the magnetogram in the spherical harmonic transform space (middle row) and the field after optimization (bottom row), as seen from three sample perspectives, $0^{\circ}$, $120^{\circ}$, and $240^{\circ}$.  Examining each column one can see substantial changes between the original and the perturbed field, which have been largely reversed in the optimized field image at the bottom.}
\label{harmonic_hairyballs}
\end{figure}
\begin{figure}
  \begin{tabular}{ c c }
    \subfloat{
	  \includegraphics[width=.45\textwidth]{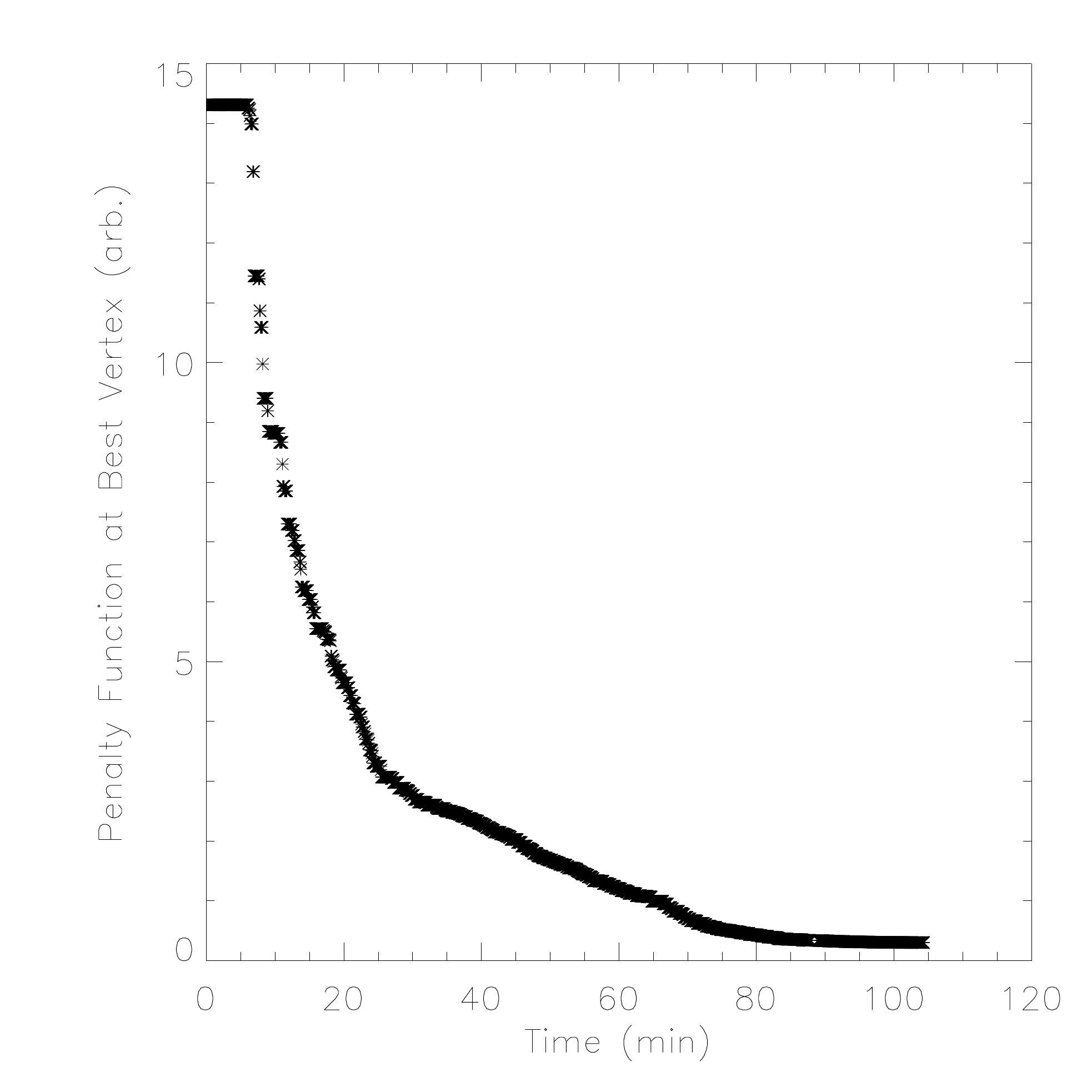}
	  \label{harmonic_convergence}} &
    \subfloat{
	  \includegraphics[width=0.45\textwidth]{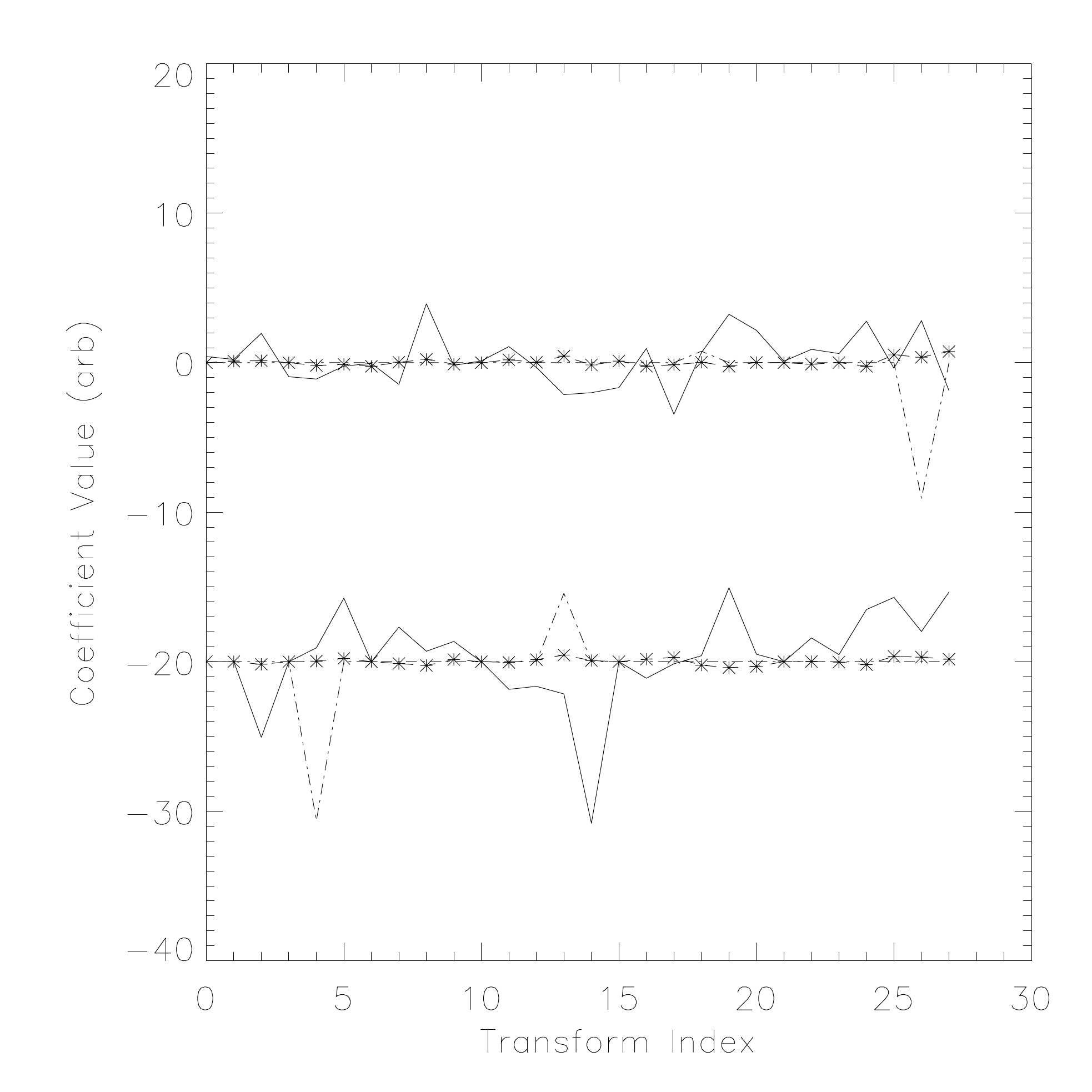}
 	  \label{harmonic_coefficients}}\\
  \end{tabular}
  \caption{(Left) Convergence of the penalty function $J$ in equation \ref{penalty} for the test problem shown in Figure \ref{harmonic_test}.  (Right) Spherical harmonic coefficients with $l < 7$ for the same test problem.  The coefficients are plotted against the transform index, which increases with $m$, then $l$.  (\emph{i.e.} index $0$ corresponds to $(l,m)=(0,0)$, index $1$ to $(l,m)=(1,0)$, index $2$ to $(l,m)=(1,1)$ and so forth.)  The upper set of curves show the real components for each coefficient, and the lower set show the imaginary components.  The solid line shows the original spherical harmonic transform coefficients, the dash-dot line represents the difference between the original and perturbed coefficients, and the dashed line with asterisks represents the difference between the original and the optimized coefficients.}
\end{figure}

 The perturbed dipole magnetogram used in the previous section is not as appropriate for testing the spherical harmonic optimization as it is for the pixel-by-pixel optimization, because the perturbation was very localized.  Additionally, while constraints from only a few well-placed image planes were sufficient for the localized feature in the perturbed dipole test above, constraining global functions like the spherical harmonics works best when the constraints are as well-resolved in longitude and latitude as the functions whose coefficients need to be optimized.  For the harmonic optimization method we created our test inputs from a \ac{gong} synoptic magnetogram from 10 January 2014.  We constructed a \ac{pfss} model based on the original magnetogram to obtain the field constraints $\{\theta_o\}$ at $8{,}712$ points spread over $11$ evenly spaced image planes, calculated the \ac{sht} of the magnetogram, and perturbed four randomly chosen coefficients.  Figure \ref{harmonic_coefficients} shows a plot of the transform coefficients before pertubation in black, overplotted to show the perturbations we added for this particular test (orange) and the residual changes after optimization (red).  For brevity, only the coefficients for $l < 7$ are shown; additional coefficients were held constant throughout the optimization.

 Figure \ref{harmonic_hairyballs} shows the original (top row), perturbed (middle), and optimized (bottom) field extrapolations from three sample perspectives, longitudes $0^{\circ}$, $120^{\circ}$ and $240^{\circ}$.  We optimized the spherical harmonic transform coefficients with $l$ equal to $1$ through $6$.  The $l=0$ coefficient, which partners with a spherical harmonic that is just an additive constant, is set to zero.  We did not need to optimize the coefficients for the imaginary part of the $m=0$ terms, as these spherical harmonics are purely real and optimizing the imaginary coefficients would be meaningless.  Additionally, we incorporated a constraint in the $O$ term of equation \ref{penalty} proportional to the net magnetic flux of the magnetogram.  With these settings the optimization takes a little over $100$ minutes (see Figure \ref{harmonic_convergence}) and produces an extrapolation that better matches the provided constraints, as can be seen in Figure \ref{harmonic_components}.  
\begin{table}
  \centering
  \renewcommand{\arraystretch}{2.1}
  \begin{tabular}{| c c |}
     \hline
    $C_{CS} \equiv \frac{1}{M} \sum \frac{\mathbf{B}_i \cdot \mathbf{b}_i}{\| \mathbf{B}_i \| \| \mathbf{b}_i \|} $  &  $C_{vec} \equiv \frac{\sum {\mathbf{B}_i \cdot  \mathbf{b}_i}}{( \sum \| \mathbf{B}_i \|^2  \sum \| \mathbf{b}_i \|^2 )^{1/2}} $ \\
    $E_n \equiv \frac{\sum \| \mathbf{b}_i -\mathbf{B}_i \|}{\sum \| \mathbf{B}_i \|}$  &  $E_m \equiv \frac{1}{M} \sum \frac{\| \mathbf{b}_i - \mathbf{B}_i \|}{\|\mathbf{B}_i \|}$ \\
   \multicolumn{2}{|c|}{ $\epsilon \equiv \frac{\Sigma \|\mathbf{b}_i \|^2}{\Sigma \| \mathbf{B}_i \|^2}$ } \\
    \hline
  \end{tabular}
  \caption{Formulas for figures of merit used to quantify the improvement due to optimization.  In all cases, $M$ is the total number of points at which the model has been calculated, $\mathbf{B_{i}}$ is the theoretical magnetic field at point $i$, and $\mathbf{b}_i$ is the magnetic field at point $i$ in our model.}
    \label{fomformulas}
\end{table}
\begin{table}
  \centering
  \begin{tabular}{| c  c  c  c  c  c  c |}
    \hline
     & MAE & $C_{cs}$ & $C_{vec}$ & $E_n$ & $E_m$ & $\epsilon$  \\
    \hline
    Perturbed & 2.05 & 0.489 & 0.984 & 0.531 & 2.43 &  1.22 \\
    Optimized  & 0.187 & 0.996 & 0.9998 & 0.0383 & .102 & 1.01 \\
    \hline
  \end{tabular}
  \caption{Values of the figures of merit defined in Table \ref{fomformulas} for the perturbed and optimized fields.}
  \label{figuresofmerit}
\end{table}
Table \ref{fomformulas} gives formulas for several figures of merit used previously by \citet{metcalf08} and \citet{schrijver06} to compare the performance of different \ac{nlfff} models on test problems.  We have applied these figures of merit to the perturbed and optimized fields of our test problem, comparing each to the true field model derived from the original \ac{gong} magnetogram, aiming to quantify the improvement seen in Figure \ref{harmonic_hairyballs}.  The values for these metrics for this test are given in Table \ref{figuresofmerit} and show a striking improvement in the model after optimization, as will be discussed further in the text below.
\section{Discussion}
\label{discussion}
 The results of testing with our model problems above show that starting from synoptic magnetograms, one can improve global coronal magnetic field models to better agree with image-based morphological constraints.  In both of the test cases presented, the optimized magnetograms and field models are significantly closer to the unperturbed magnetograms and fields on which the constraints were based. 

\subsection{Perturbed Dipole Test}
In the perturbed dipole test the optimization method was able to largely neutralize a monopolar perturbation and return the magnetic field model to a dipolar configuration, as can be seen by comparing the left and right hemispheres of Figure \ref{dipole_output_hairyball} after the optimization.  Not only is there clear visual improvement in the field morphology, but there is quantifiable improvement in the magnetogram as well; as shown in Figure \ref{dipole_mae}, the \ac{mae} of the perturbed region falls from $0.324$ to $0.211$ in the first five minutes of the optimization.  These results were acheived with a minimum number of input constraints, taken from only two image planes, evenly distributed in longitude.  The optimization process has introduced some noise into the magnetogram, but only to the extent that it did not interfere with the agreement between the model field and the observed field morphology.  There must be a net improvement (decrease) in the value of the penalty function, $J$.  Solutions which do not result in a net improvement will be rejected.

The perturbed dipole test is intriguing because it suggests that the proposed technique may be able to help alleviate a problem that has plagued coronal modelers for some time: the lack of observations of emergence, dissipation, or cancellation of magnetic flux on the far side of the sun.  In the perturbed dipole test we have a magnetogram flaw placed near the limb of the sun that would not have been corrected in the synoptic magnetogram for at least another two days, but whose presence would have had a noticeable effect on the field morphology far into the corona.  If the coronal magnetic field model were naively constructed based on this flawed magnetogram, its effects would have carried over into models of the heliospheric magnetic field, the solar wind, and \ac{cme} propagation.  Meanwhile, the true coronal field morphology would already have been apparent to coronagraphs on the Sun-Earth line.  Our results suggest that optimization would help resolve this problem.

\subsection{Perturbed Transform Test}
In the second test we have presented, the effect of the perturbations on individual pixels is small, but covers the entire image.  The first and second rows of Figure \ref{harmonic_hairyballs} show the change in the model field induced by the perturbation from three evenly spaced viewpoints, giving the original field in the top row and the perturbed field in the middle.  Comparing these to the bottom row - after optimization - one can see that much of the original structure has been restored.  Spurious regions of open flux have been removed, including a high-latitude band-like structure seen in the northern hemisphere in the third column and a large coronal hole in the center column.  Also, a southern hemisphere streamer belt seen in the left-hand column, introduced by the perturbation, has been removed during optimization.  Figure \ref{harmonic_components} shows that optimization also produced a clear improvement in agreement between the field model and our constraints.  

Looking at Table \ref{figuresofmerit} one can see that each of our figures of merit has been improved by the optimization, quantifying in several ways the correction seen in Figure \ref{harmonic_hairyballs}.   $C_{CS}$ and $C_{vec}$ both largely measure angular differences between the correct and the model fields, and have ideal values of $0$.  It is perhaps not too surprising that they have been reduced by the optimization; $C_{vec}$ was nearly ideal even for the perturbed field, and these metrics largely measure something we have deliberately set out to minimize; what is surprising is that the $E_n$, $E_m$, and $\epsilon$ metrics  (all with ideal values of $1$) also show improvement.  These metrics are sensitive to differences in both vector direction and norm, something we have not directly constrained.  Also, though our primary goal is to optimize the field model, we found that if we neglect the effects of transforming and inverse transforming (thereby losing some of the higher frequency information), the \ac{mae} of the magnetogram is also reduced by the optimization.  We found the \ac{mae} of the perturbed magnetogram to be $2.05$, and after the optimization to be $0.187$.  

Figure \ref{harmonic_coefficients} shows the size of the perturbations in the transform space before and after optimization.  Here one can see that the introduced perturbations are comparable in magnitude to the original coefficient values prior to optimization, and that after optimization the perturbations have been largely eliminated.  The largest of the remaining problematic points after optimization tend to be associated with high-frequency harmonics (high $m$ or $l$ value).

In contrast to the perturbed dipole test, here the test magnetogram is not obviously in error.  Visual inspection of the magnetogram before and after perturbation reveals no noticeable changes, as the average absolute change in magnetogram flux due to the perturbations in the transform space is only about $2$ Gauss, which is quite small compared to a magnetogram dynamic range of $1100$ Gauss.  (Note that again we are neglecting changes due to the spherical harmonic transformation process itself, which tends to diminish regions of strong flux.)  This suggests that the presented optimization technique can be used to overcome subtle magnetogram flaws that may result, for example, from being unable to view one of the polar regions, or from neglecting the effects of differential rotation.

\section{Conclusions and Future Work} 
\label{conclusions}
There is a very real need for fast, global models of the coronal magnetic field in the solar physics community.  The continued prevalence of the \ac{pfss} model in the literature despite its well-known faults is a striking testimony to that need.  What we have described here is a method for improving such global models by utilizing the morphological constraints that can be acquired from coronal images.  We have tested this method on simple model problems with known solutions, using minimal computational resources.

The results of the test presented here show that optimization with such morphological constraints works reasonably well with both high-amplitude localized perturbations (mimicking the emergence or cancellation of a region of concentrated flux on the far side of the sun), and for low-amplitude, non-localized perturbations such as may be commonly found on synoptic magnetograms due to unobserved polar regions or surface motions not accounted for in the synoptic magnetogram creation.

Having done this initial testing, several obvious next steps present themselves.  First, in the spherical harmonic transform version of the optimization software we have limited the optimization to a subset of the transform coefficients.  A similar restriction my be practicable for the pixel-by-pixel optimization, but needs further investigation.  

Additionally, in the interest of further validation, we need to conduct a comparison of naive and optimized field models with the limited available coronal magnetic field measurements.  This requires the creation of high-resolution coronal models optimized with solar images in the range $1-2.5 R_\sun$, such as those from the \ac{mlso} K-Cor instrument.  

We would also like to apply this technique to other coronal models.  As we have mentioned before, there is no reason the underlying model we optimize needs to be potential - image-based constraints can be used to improve any coronal model, and this method can and should be applied to other types of models.  

Once this technique has been perfected and fully validated, it will provide better coronal models, which can provide the lower boundary condition for heliospheric models. With improved coronal models we can better predict solar wind speed and the \ac{imf}, a topic of particular interest due to the upcoming \ac{so} and \ac{spp} missions.
\section{Acknowledgements}
We wish to thank Marc DeRosa for contributing the \ac{pfss} library to SolarSoft, and for helpful reponses to questions about the software.  We also wish to acknowledge Figen Oktem for useful discussions regarding regularization in optimization problems, and Leon Ofman for similarly useful discussions of coronal \ac{mhd} simulations.

\bibliography{master_extrapolation}

\end{document}